\documentclass{nature_plusfigure}
\usepackage{graphicx}
\usepackage{aas_macros,amsmath,amssymb,xcolor,url}
\usepackage{multirow}
\usepackage{longtable}
\graphicspath{{./}{figures/}}

\usepackage[symbol]{footmisc}

\usepackage{cleveref}

\usepackage[none]{hyphenat}
\usepackage{multibib}
\newcites{New}{References}
\usepackage{lineno}
\usepackage{subcaption} 

\title{Optical follow-up of the neutron star-black hole mergers S200105ae and S200115j}

\author{Shreya Anand$^{1}$\footnote[1]{\label{authorship} These two authors contributed equally to this work.},
	Michael W. Coughlin$^{1, 2*}$,
	Mansi M. Kasliwal$^{1}$,
	Mattia Bulla$^{3}$,
	Tom{\'a}s Ahumada$^{4}$,
	Ana Sagu{\'e}s Carracedo$^{5}$,
	Mouza Almualla$^{6}$,
	Igor Andreoni$^{1}$,
	Robert Stein$^{7,8}$,
	Francois Foucart$^{9}$,
	Leo P. Singer$^{10,11}$,
	Jesper Sollerman$^{12}$,
	Eric C. Bellm$^{13}$,
	Bryce Bolin$^{1}$,
	M. D. Caballero-Garc\'ia$^{14}$,
	Alberto J. Castro-Tirado$^{15,16}$,
	S. Bradley Cenko$^{10,11}$,
	Kishalay De$^{1}$,
	Richard G. Dekany$^{17}$,
	Dmitry A. Duev$^{1}$,
	Michael Feeney$^{17}$,
	Christoffer Fremling$^{1}$,
	Daniel~A.~Goldstein$^{1}$,
	V. Zach Golkhou$^{13,18}$,
	Matthew J. Graham$^{1}$,
	Nidhal Guessoum$^{6}$,
	Matthew J. Hankins$^{1}$,
	Youdong Hu$^{15,19}$,
	Albert K. H. Kong$^{20}$,
	Erik C. Kool$^{12}$,
	S. R. Kulkarni$^{1}$,
	Harsh Kumar$^{21}$,
	Russ R. Laher$^{22}$,
	Frank J. Masci$^{22}$,
    Przemek Mr{\'o}z$^{1}$,
    Samaya Nissanke$^{23}$
    Michael Porter$^{17}$,
    Simeon Reusch$^{7,8}$,
    Reed Riddle$^{17}$,
	Philippe Rosnet$^{24}$,
	Ben Rusholme$^{22}$,
	Eugene Serabyn$^{25}$,
	R. S\'anchez-Ram\'irez$^{26}$,
    Mickael Rigault$^{24}$,
    David L. Shupe$^{22}$,
    Roger Smith$^{17}$,
    Maayane T. Soumagnac$^{27,28}$,
    Richard Walters$^{17}$
    and Azamat F. Valeev$^{29}$
	}

\newcommand{\arcsec}{$^{\prime\prime}$}

\begin{document}

\maketitle

\begin{affiliations}
\item{Division of Physics, Mathematics, and Astronomy, California Institute of Technology, Pasadena, CA 91125, USA}

\item{School of Physics and Astronomy, University of Minnesota, Minneapolis, Minnesota 55455, USA}


\item{Nordita, KTH Royal Institute of Technology and Stockholm University, Roslagstullsbacken 23, SE-106 91 Stockholm, Sweden}

\item{Department of Astronomy, University of Maryland, College Park, MD 20742, USA}

\item{The Oskar Klein Centre, Department of Physics, Stockholm University, AlbaNova, SE-106 91 Stockholm, Sweden}

\item{American University of Sharjah, Physics Department, PO Box 26666, Sharjah, UAE}


\item{Deutsches Elektronen Synchrotron DESY, Platanenallee 6, 15738 Zeuthen, Germany}
\item{Institut f{\"u}r Physik, Humboldt-Universit{\"a}t zu Berlin, D-12489 Berlin, Germany}

\item{Department of Physics, University of New Hampshire, 9 Library Way, Durham NH 03824, USA}

\item{Astrophysics Science Division, NASA Goddard Space Flight Center, MC 661, Greenbelt, MD 20771, USA}
\item{Joint Space-Science Institute, University of Maryland, College Park, MD 20742, USA}

\item{The Oskar Klein Centre, Department of Astronomy, Stockholm University, AlbaNova, SE-106 91 Stockholm, Sweden}

\item{DIRAC Institute, Department of Astronomy, University of Washington, 3910 15th Avenue NE, Seattle, WA 98195, USA}

\item{Astronomical Institute of the Academy of Sciences, Bocn\'i II 1401, CZ-14100 Praha 4, Czech Republic.}

\item{Instituto de Astrof\'isica de Andaluc\'ia (IAA-CSIC), Glorieta de la Astronom\'ia s/n, E-18008, Granada, Spain}
\item{ Departamento de Ingenier\'ia de Sistemas y Autom\'atica, Escuela de Ingenieros Industriales, Universidad de M\'alaga, Unidad Asociada al CSIC, C. Dr. Ortiz Ramos sn, 29071 M\'alaga, Spain}



\item{Caltech Optical Observatories, California Institute of Technology, Pasadena, CA 91125, USA}





\item{The eScience Institute, University of Washington, Seattle, WA 98195, USA}




\item{Universidad de Granada, Facultad de Ciencias Campus Fuentenueva S/N CP 18071 Granada, Spain}

\item{Institute of Astronomy, National Tsing Hua University, Hsinchu 30013, Taiwan}



\item{Indian Institute of Technology Bombay, Powai, Mumbai 400076, India}

\item{IPAC, California Institute of Technology, 1200 E. California Blvd, Pasadena, CA 91125, USA}






\item{Center of Excellence in Gravitation and Astroparticle Physics, University of Amsterdam, Netherlands}
\item{Universit\'e Clermont Auvergne, CNRS/IN2P3, Laboratoire de Physique de Clermont, F-63000 Clermont-Ferrand, France}


\item{Jet Propulsion Laboratory, California Institute of Technology, Pasadena, CA 91109, USA}

\item{INAF - Instituto di Astrofisica e Planetologia Spaziali, Via Fosso del Cavaliere 100, 00133 Roma, Italy.}




\item{Lawrence Berkeley National Laboratory, 1 Cyclotron Road, Berkeley, CA 94720, USA}
\item{Department of Particle Physics and Astrophysics, Weizmann Institute of Science, Rehovot 76100, Israel}


\item{Special Astrophysical Observatory, Russian Academy of Sciences, Nizhnii Arkhyz, 369167 Russia}
\end{affiliations}

\begin{abstract}
LIGO and Virgo’s third observing run (O3) revealed the first neutron star–black hole (NSBH) merger candidates in gravitational waves. These events are predicted to synthesize r-process elements\cite{LaSc1974,LiPa1998} creating optical/near-IR “kilonova” (KN) emission. The joint gravitational-wave (GW) and electromagnetic detection of an NSBH merger could be used to constrain the equation of state of dense nuclear matter\cite{CoDi2018}, and independently measure the local expansion rate of the universe\cite{Schutz1986}. Here, we present the optical follow-up and analysis of two of the only three high-significance NSBH merger candidates detected to date, S200105ae and S200115j, with the Zwicky Transient Facility\cite{Bellm2018} (ZTF). ZTF observed $\sim$\,48\% of S200105ae and $\sim$\,22\% of S200115j’s localization probabilities, with observations sensitive to KNe brighter than $-$17.5\,mag fading at 0.5\,mag/day in g- and r-bands; extensive searches and systematic follow-up of candidates did not yield a
viable counterpart. We present state-of-the-art KN models tailored to NSBH systems that place constraints on the ejecta properties of these NSBH mergers. We show that with depths of $\rm m_{\rm AB}\approx 22$ mag, attainable in meter-class, wide field-of-view survey instruments, strong constraints on ejecta mass are possible, with the potential to rule out low mass ratios, high BH spins, and large neutron star radii.

\end{abstract}

During O3, LIGO and Virgo detected eight NSBH and six BNS candidate events at various confidence levels, with localization regions spanning a few tens to several thousands of square degrees and median distances in the range $\sim$108-630\,Mpc. We do not include S190718a as a BNS merger candidate due to glitches in the detectors near trigger time, which have a very high terrestrial probability ($>$ 98\%).
All of the NSBH candidates had $\sim$100\% probability of one of the component masses being $<$\,3 $M_\odot$, and therefore likely to be a neutron star. 
Only two candidates, S200105ae\cite{gcn26640} and S200115j\cite{gcn26759}, initially had finite probability of leaving behind a non-zero amount of neutron star material outside the final black hole, although S200115j's updated analysis \cite{gcn26807} gives $<1$\% probability of leaving behind a remnant. 
S200105ae\cite{gcn26640} and S200115j\cite{gcn26759} were both detected in January, at 2020-01-05 16:24:26.057 and 2020-01-15 04:23:09.742 UTC respectively (see Methods). During O3, ZTF ran a dedicated follow-up program to identify optical counterparts to gravitational-wave (GW) candidates (e.g. Ref~\cite{CoAh2019b,gcn24237,gcn25324}). Together with the Global Relay of Observatories Watching Transients Happen (GROWTH) network (http://growth.caltech.edu/), ZTF rapidly followed up and classified objects that were consistent with the candidates.
Over the 3 nights following detection, ZTF covered 3300\,deg$^2$ and 1100\,deg$^2$ for S200105ae and S200115j respectively, corresponding to $\sim$\,52\% of the localization probability for S200105ae, and $\sim$22\% of the localization probability for S200115j (see Methods).  S200115j occurred during Palomar nighttime, so our triggered observations began immediately, but poor weather on the two nights following the merger prevented further follow-up observations.

\begin{figure*}[t]
 \centering
\includegraphics[width=4.5in]{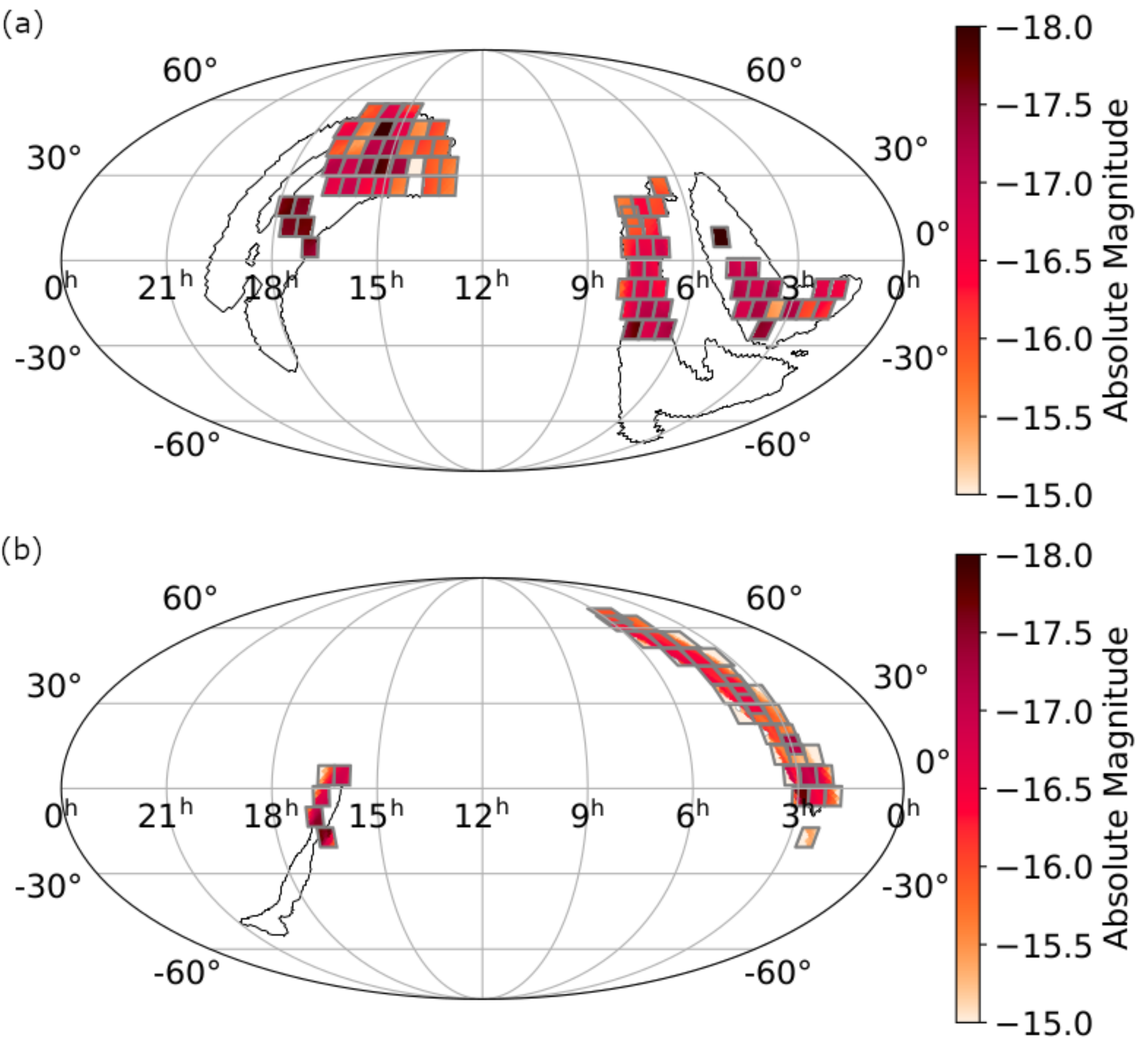}
  \caption{\textbf{Absolute magnitudes corresponding to ZTF pointings in the skymap.} We map the absolute magnitudes corresponding to the distance provided in the GW LALInference skymap, measured at the center of each field, and the deepest limiting magnitude in either $g$- or $r$-bands (computed as a median over the CCDs in a particular field) for S200105ae (a) and S200115j (b). We also show the 90\% probability region contours to guide the eye.}
 \label{fig:absolute}
\end{figure*}

As a metric for understanding the efficacy of ZTF's observations, we show the mean absolute magnitude to which we are sensitive as a function of sky location in Figure~\ref{fig:absolute}. This folds in the distance distribution across the skymap compared to our median limiting magnitude in each of the fields (See \ref{fig:limmag}). The best limiting magnitudes correspond to absolute magnitudes $\lesssim -16$ mag for both events, with typical observations ranging from M$\sim\,-16.5$ mag to M$\sim\,-17.5$ mag. AT2017gfo~\cite{2017Sci...358.1556C}, the optical counterpart to GW170817, peaked at M$\sim-16$ mag, and KNe from NSBH models are typically brighter than those from BNSs~\cite{RoFe2017,Barbieri:2019bdq,Kawaguchi2020}, indicating that our observations are in the magnitude range required for detection.

\begin{figure*}[t]
  \includegraphics[width=1\textwidth]{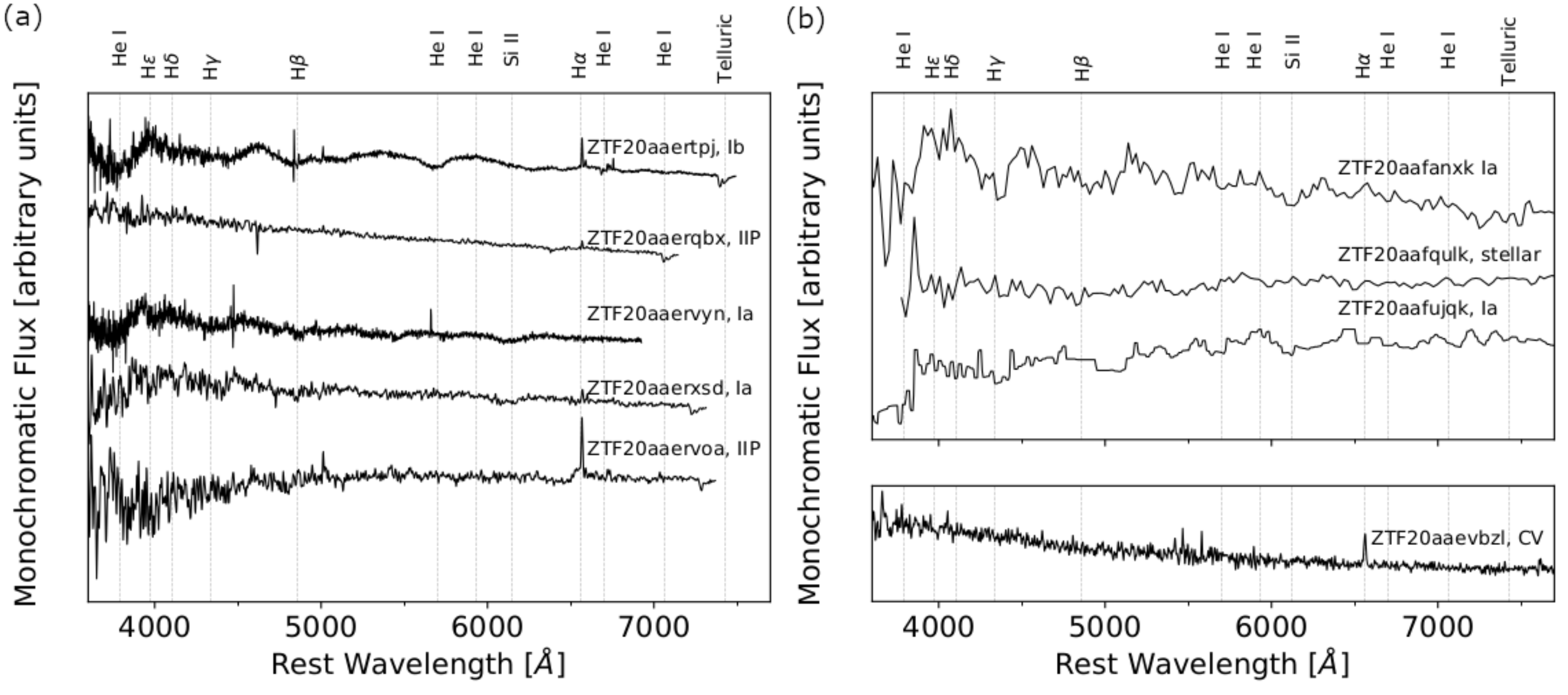}
  \caption{\textbf{Spectra of all of the candidates ruled out spectroscopically during both campaigns.} In order to visualize all the spectra on the same figure, we have applied a vertical offset to the flux, and plotted each spectrum at mean signal-to-noise ratio. The vertical dashed lines correspond to common spectral absorption and emission features in SN spectra. (a) Spectra of five S200105ae candidates taken with the Optical System for Imaging and low Resolution Integrated Spectroscopy (OSIRIS) on the Gran Telescopio Canarias (GTC) of the Roque de los Muchachos Observatory in La Palma, Spain \cite{gcn26702, gcn26703}.  The top three spectra were taken on Jan 11th, and the bottom two were taken on Jan 10th. From top to bottom, ZTF20aaertpj was classified as a SN Ib at z(s) = 0.026, ZTF20aaerqbx was classified as a SN IIP at z(s) = 0.098, ZTF20aaervyn was shown to be a SN Ia at z(s) = 0.112, ZTF20aaerxsd is a SN Ia at z(s) = 0.055, and ZTF20aaervoa was classified as a SN IIP at z(s) = 0.046. (b) Top: all spectra taken with the SED Machine (SEDM) on the Palomar 60-inch telescope (P60); from top to bottom, ZTF20aafanxk (S200105ae) was classified as a SN Ia at z(s) = 0.103 on January 18th, the spectrum of ZTF20aafqulk (S200115j), observed on January 24th, indicates that it is likely stellar, and ZTF20aafujqk (S200105ae), also observed on January 18th, was classified as a SN Ia at z(s) = 0.074. (b) Bottom: The spectrum of ZTF20aaevbzl (S200105ae) taken by the Double Spectrograph (DBSP) on the Palomar 200-inch telescope (P200) obtained on January 18th, 2020, contains a H$\alpha$ feature in a mostly featureless blue continuum that is indicative of it being a cataclysmic variable.
  }
 \label{fig:spectra}
\end{figure*}

In addition to requiring multi-epoch coverage of large localizations at sufficient depth, these searches normally yield hundreds of thousands of alerts that require quick and thorough vetting (see Methods for specific criteria and \ref{fig:table_alert}). We successfully narrowed this list down to a select few candidates consistent with our criteria within minutes for both events; only 22 candidates for S200105ae and 6 candidates of S200115j remained (see Methods for selection criteria).
GROWTH obtained follow-up photometry and spectroscopy for the candidates passing our requirements to assess their relation to either event. 
Using a global array of telescopes (see Methods for observatories and instruments),  we reject each of our candidates based on the following criteria:
\begin{itemize}
    \item Spectroscopic Classification: candidates spectroscopically determined to be supernovae or other transient (see Figure \ref{fig:spectra} and \ref{fig:lc_transients}).
    \item Slow photometric evolution: candidates evolving at $<|0.3|$ mag/day, below the expected fast evolution for KNe over the course of a week (see Methods and \ref{fig:evolution} for justification and \ref{fig:S200105ae_lcs} for candidate lightcurves).
    \item Stellar Variables: candidates coincident with point sources, likely to be variable stars or cataclysmic variables in the Milky Way.
    \item Slow-moving asteroids: candidates that are later determined to be asteroids or other solar-system objects (see \ref{fig:ast_ZTF20aaegqfp}).
\end{itemize}
After thorough vetting, we found no candidate remaining that could plausibly be associated with either event (see the candidates spatial distribution in \ref{fig:skymap} and the list of the candidates in Supplementary Information ~\cref{table:S200105ae_spec_followup,table:S200105ae_followup,table:S200115j_followup}).

\begin{figure*}[t]
\includegraphics[width=1\textwidth]{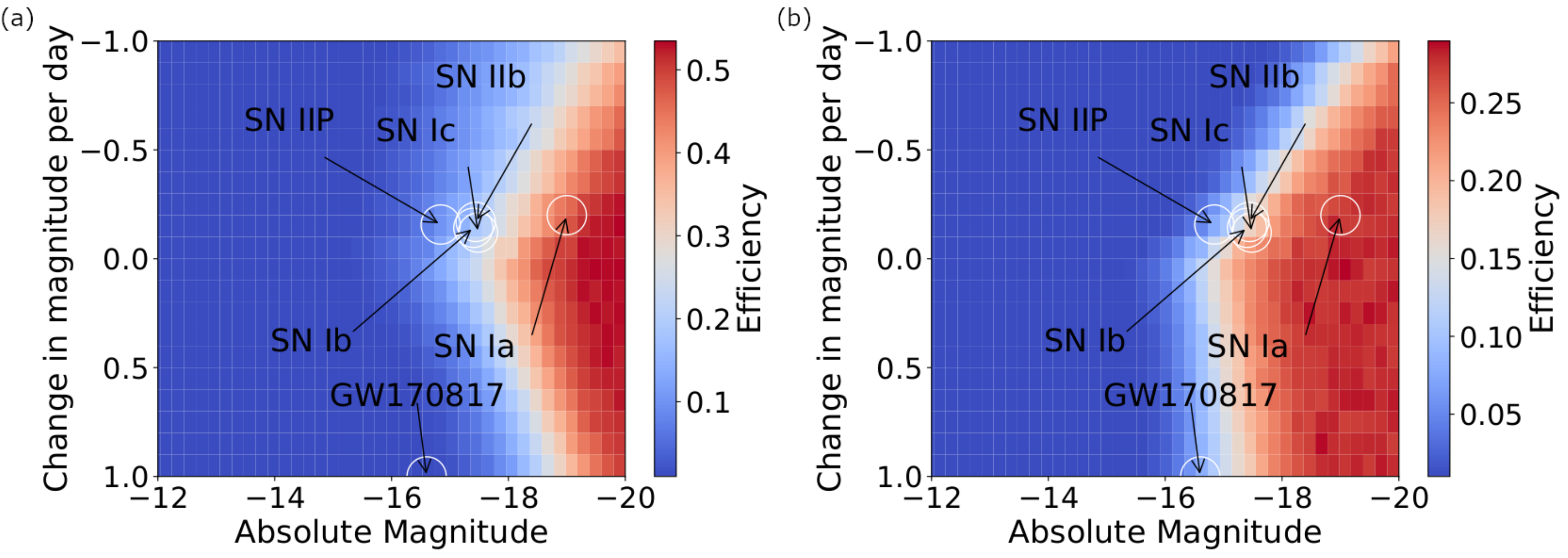}
  \caption{\textbf{Detection efficiency of simulated KNe based on ZTF observations.} Ratio of recovered vs injected KNe (efficiency) identified in observations in a skymap for an analytic model varying absolute magnitude and change in magnitude per day for (a) S200105ae and (b) S200115j in both $g$ and $r$-bands. Here, the magnitude corresponds to the peak absolute magnitude of the injected kilonovae for a linear model with a given rise or decay rate. The maximum of the colorbar scale is set to the maximum efficiency achieved (at M$=-20$), which for S200105ae was 53\% and 29\% for S200115j. We include approximate peak absolute magnitudes and approximate rise rates for some common SNe types; for GW170817, we plot the absolute magnitude at detection and the approximate decline rate to guide the eye.}
 \label{fig:efficiency}
\end{figure*}

\begin{figure*}[t]
 \begin{center}
 \includegraphics[width=1\textwidth]{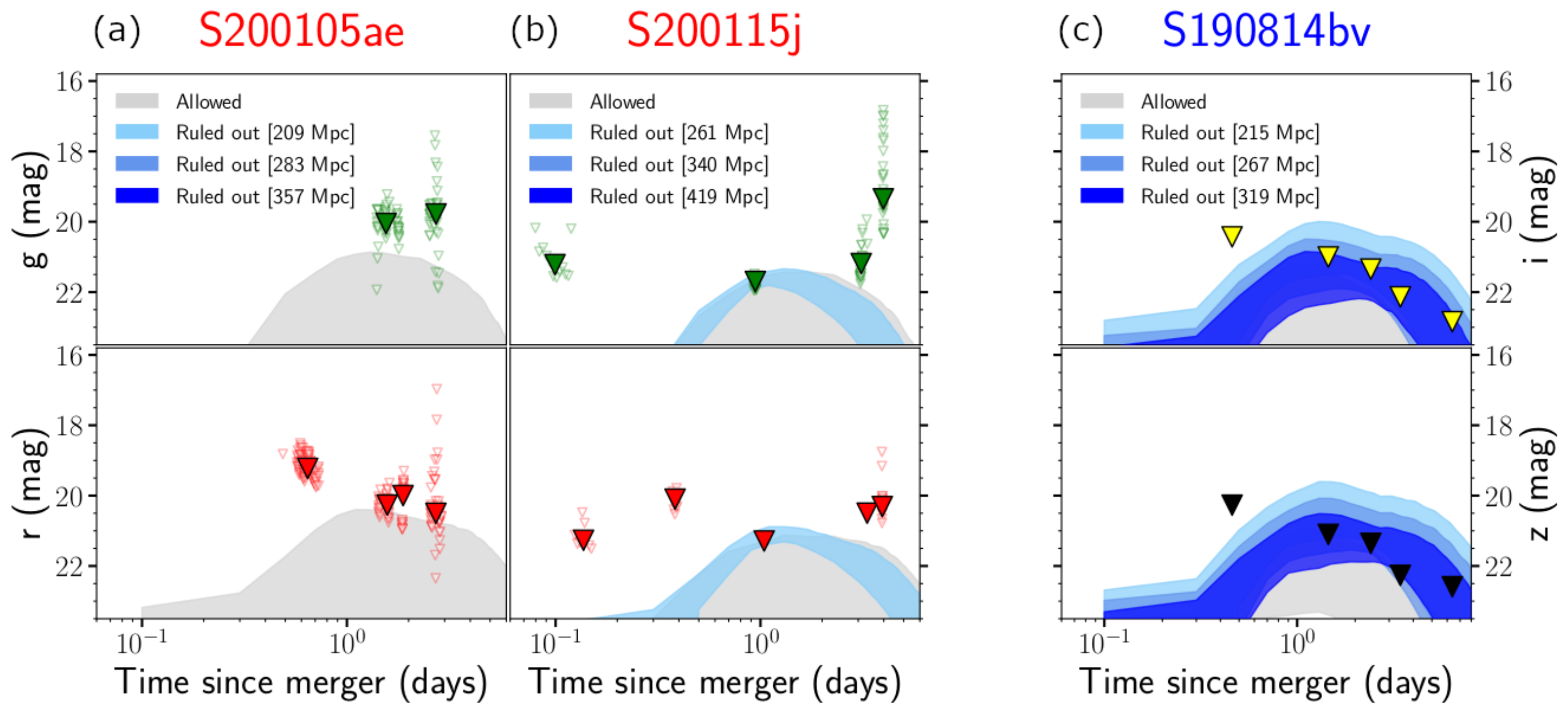}
  \caption{\textbf{Constraints on kilonova model parameters based on median limiting magnitudes.} We display all KN light curves ruled out by \textit{median} 5-$\sigma$ limits on (a) S200105ae (ZTF), (b) S200115j (ZTF) and (c) S190814bv (DECam). For S200105ae and S200115j, median AB magnitudes are shown with filled triangles, while individual limits are shown with open triangles. Limits shown for S190814bv are median depth values from table 1 of Ref.~\cite{Andreoni2020}. KN light curves are calculated with \textsc{possis} \protect\cite{Bulla2019}; we show in blue when they are ruled out by the limits at three different distances (corresponding to median distances and $\pm$1$\sigma$ distance uncertainties from LIGO) and in grey otherwise. For each distance, the shaded area represents the range spanned by different models and different viewing angles (with the brighter end generally corresponding to higher masses and polar orientations while the fainter end to lower masses and equatorial orientations). The median limits for S200105ae do not constrain any kilonova models for any distance assumptions, while for S200115j they place constraints only on the models for nearby kilonovae (light blue). For S190814bv, median limits constrain kilonova models for all distance assumptions.}
 \label{fig:kn_lc}
 \end{center}
\end{figure*}

The non-detection in our searches allows us to impose both empirical and model-based constraints on photometric evolution for a counterpart falling within the observed region. To place the coverage and limits in context, we compare our observations to empirical models of evolution with a linear rise and decay (Figure~\ref{fig:efficiency}), and KN models, which allow ejecta masses to vary (Figure~\ref{fig:kn_lc}). 
Using \texttt{simsurvey}\cite{FeNo2019} to inject and recover simulated KNe, we show in Figure~\ref{fig:efficiency} that ZTF should have detected a KN in the observed region of either skymap brighter than M\,$\lesssim$\,-17.5\,mag and fading slower than $0.5$\,mag per day in both $g$ and $r$-bands. We simulate kilonovae with various absolute magnitudes and evolution rates assuming no color evolution. Our recovery criteria requires a single kilonova detection in either filter. We plot the KN absolute magnitudes at peak along with their evolution rates. We also mark AT2017gfo, which had a peak absolute magnitude of about $-16$\,mag in optical bands, fading at $\sim 1.0$\,mag per day in $g$- and $r$-bands.  Lack of observations on the first night for S200105ae, owing to a delay in the release of the initial skymap, worsened constraints compared to S200115j (see Methods).  We note here that our sensitivity to rising or fading kilonovae is highly dependent on latency in starting observations and number of follow-up epochs. 

For our model-dependent constraints, assuming the kilonova is in the area observed, we take a series of representative median magnitudes for each night of observations and compare them to lightcurve models from the radiative transfer code \textsc{possis} \cite{Bulla2019}; we generated them using a new grid of KN spectra tailored to NSBH mergers. These are summarized in Figure~\ref{fig:kn_lc}, where we show light curves that are allowed (grey) or ruled out at different distances (light to dark blue) by the median magnitudes achieved with our observations of S200105ae and S200115j (see Methods). We find that the median magnitudes place weak constraints on these models. Specifically, all KN light curves we consider are fainter than the limits for S200105ae while only a few models with large amounts of post-merger ejecta ($\gtrsim0.05 M_\odot$) are ruled out for S200115j at polar viewing angles and for the nearest-by portions of the skymap. Additionally, we note that due to our coverage in both skymaps being less than 50\%, our model constraints for S200105ae and S200115j only apply within the observed region.  
For comparison, the right panel of Figure~\ref{fig:kn_lc} shows NSBH models from our new grid that are ruled out by the DECam observations of S190814bv~\cite{Andreoni2020}; such limits are more robust than our limits on S200105ae and S200115j due to DECam covering 98\% of the skymap (compared to 48\% and 22\%). For that well-localized event, the deeper DECam limits and the closer distance for S190814bv (d=267$\pm$52 Mpc [ref.\cite{gcn25324}]) lead to a larger number of models ruled out.


To understand the scientific performance and potential of meter-class, wide field-of-view imagers as powerful tools in EM-GW follow-up, we determine what constraints are possible on the viewing angle of a potential counterpart, the dynamical ($M_\mathrm{ej,dyn}$) and post-merger ($M_\mathrm{ej,pm}$) ejecta and the binary parameters with the deepest ZTF exposures on each night (see Methods). For S200105ae, with five-minute exposures reaching a depth of $\rm m_{\rm AB}\gtrsim22$ mag, ZTF would be sensitive to a large fraction of KNe with polar and intermediate viewing angles. Non-detection of a kilonova in these circumstances could rule out $M_\mathrm{ej,dyn}\leq0.02\,M_\odot$ and $M_\mathrm{ej,pm}\leq0.04\,M_\odot$ for polar directions at 283 Mpc (see \ref{fig:kne}).  Using these $M_\mathrm{ej}$--$\theta_\mathrm{obs}$ constraints, we could estimate the maximum aligned spin of the BH component for different assumptions on the viewing angle, binary mass ratio and neutron star radius. Non-detection would further rule out low mass ratios, high BH spins, and/or large neutron star radii (see \ref{fig:pe}). For high mass ratios, the limit on $M_{\rm ej,dyn}$ would be more constraining than the limit on $M_{\rm ej,pm}$. As $M_{\rm ej,dyn}$ is reasonably well known from simulations~\cite{Foucart:2018rjc}, our modeling of the ejected mass is not a significant source of uncertainty. For low mass ratios, the limit on $M_{\rm ej,pm}$ would be more constraining. Current simulations only allow us to constrain $M_{\rm ej,pm}$ to within a factor of $2-3$ [ref.~\cite{Christie:2019lim}], and are in this case an important source of modeling uncertainty. Here, we derive an upper limit on the black hole spin using a conservative estimate of $M_{\rm ej,pm}$. Improved simulations providing better estimates of $M_{\rm ej,pm}$ could make these limits more constraining in the future (see \ref{fig:pelim} for the binary parameter region not constrained by our simulations). 

Additionally, the available parameter space could be significantly reduced if we knew the chirp mass of the binary~\cite{Barbieri:2019bdq}, which is not yet published by LIGO-Virgo.
For S200115j, whose median distance was $\sim$ 60\,Mpc greater than S200105ae, the deepest exposures would only be sensitive to kilonovae at nearby distances, and thus place weak constraints on the binary parameters.

Revisiting the follow-ups of S190814bv with the updated NSBH grid, we find more stringent constraints on the ejecta mass and binary parameters than for S200105ae, even using median observations (Figure~\ref{fig:kn_lc}). Polar orientations are ruled out at distances $\leq267$ Mpc, limiting the ejecta masses to $M_\mathrm{ej,dyn}\lesssim0.01\,M_\odot$ and $M_\mathrm{ej,pm}\lesssim0.01\,M_\odot$. At intermediate orientations ($46^\circ\lesssim\theta_\mathrm{obs}\lesssim53^\circ$), these constraints are still $M_\mathrm{ej,dyn}\lesssim0.02\,M_\odot$ and $M_\mathrm{ej,pm}\lesssim0.03\,M_\odot$ (see \ref{fig:kne}). We also
find that deep $i$- and $z$-band exposures contribute significantly towards constraining a larger portion of the M$_\mathrm{ej}$--$\theta_\mathrm{obs}$ and binary parameter-space (see \ref{fig:pe2}). Literature on kilonova models\cite{TaHo2014, Kawaguchi2020} have predicted kilonovae from NSBH mergers to be brighter in the $i$- and $z$-bands compared to $g$- and $r$-bands. The same reddened emission is evident in our models (see \ref{fig:nsbh_lc} and \ref{fig:iz_vs_gr}), and is demonstrated by our re-analysis of the DECam upper limits on S190814bv. Thus observations in redder bands will yield better overall constraints on NSBH kilonova emission.

Several works in the literature\cite{RoFe2017,Barbieri:2019bdq,Kawaguchi2020} have shown that KNe from NSBH mergers are generally brighter than those resulting from BNS mergers. A similar behaviour is found in NSBH and BNS models computed here and in Ref.~\cite{Dietrich2020}, respectively. Although the comparison is sensitive to the specific binary properties and thus ejecta masses adopted, we identify some general behaviour using typical values from analytical models calibrated to numerical simulations~\cite{Foucart:2018rjc,Kruger:2020gig} (e.g. for a $1.2M_\odot - 1.4M_\odot$ BNS merger with $R=12\,{\rm km}$: $M^\mathrm{BNS}_\mathrm{ej,dyn}=0.005\,M_\odot$, $M^\mathrm{BNS}_\mathrm{ej,pm}=0.05\,M_\odot$; for a $1.2M_\odot - 6M_\odot$ NSBH merger with BH spin of 0.75: $M^\mathrm{NSBH}_\mathrm{ej,dyn}=0.05\,M_\odot$ and $M^\mathrm{NSBH}_\mathrm{ej,pm}=0.05\,M_\odot$). At peak, the difference in brightness between NSBH and BNS mergers is relatively small in both $g$- and $r$-bands. The evolution after peak, however, is significantly different between the two systems. Compared to BNS mergers, NSBH mergers produce $\sim$10$\times$ more massive dynamical ejecta and are thus associated with longer diffusion timescales, as photons take longer to diffuse out of the high-density and lanthanide-rich dynamical ejecta. Consequently, KNe from NSBH mergers evolve more slowly after peak and therefore stay bright longer than those resulting from BNS mergers. The difference can be as large as $\Delta m\sim$~2 mag about 3~days post-peak for favourable viewing angles. The different evolution post-peak explains why constraints derived above for S190814bv are tighter than those using BNS models~\cite{Andreoni2020}. The slower evolution of NSBH compared to BNS mergers makes the former promising candidates for future follow-up studies. This slower evolution is fairly robust to the choice of parameters as long as the NS is disrupted by its BH companion.

Looking forward, achieving increased and consistent depth over our observations, and supplementing $r$- and $g$-band observations with an $i$-band observation will be key to increasing our chances of finding a kilonova and/or discerning properties of the merger (See Methods). NSBH binaries, with a combination of intrinsically longer-lasting emission, higher signal-to-noise ratios and therefore smaller sky areas (sky area $\sim\frac{1}{\mathrm{SNR}^2}$), and high rates based on the three high-significance NSBH candidates observed during O3 makes them ideal for counterpart searches, important for measuring the Hubble Constant given their improved inclination measurements over BNS counterparts\cite{ViCh2018}.  Furthermore, the uncertainty over the time delay between a merger and its peak lightcurve motivates obtaining observations one night after the merger; the most constraining limits from our analysis correspond to one night post-merger, when the KN is brightest (see Figure \ref{fig:kn_lc}). While low-latency follow-up is crucial for determining whether an early-time lanthanide-free component is present in these KNe, observations one night after are equally important for detection or placing ejecta mass constraints.  In this work, we have showcased a novel methodology for deriving significant constraints on NSBH kilonova models even in the case of non-detection of a counterpart, and demonstrated that such valuable constraints are within reach of wide field-of-view, meter-class imagers.

To close, we highlight the immense promise of undertaking searches for the kilonova counterparts of NSBH mergers.
The dearth of electromagnetic observations of NSBH systems as compared to BNS systems (discovered in X-ray binaries), and the difficulty of distinguishing between a low-mass BBH and a NSBH system from the GWs points to the ``smoking gun" nature of KNe in confirming the existence of such systems.
KNe are amongst the most valuable probes of the empirical ``mass gap'' between the stellar mass neutron star and black hole systems, and will allow us to observationally confirm the correlation between the mass ratio of the binary and the fate of the remnant, even in the case of non-detection. These could be jointly addressed by GW and EM facilities that possess a combination of large fields-of-view and deep sensitivity.
Continuing follow-ups of NSBH mergers is essential in granting key insights into the nature of the elusive NSBH population as a whole.

\newpage

\clearpage
\bibliographystyle{naturemag}
\bibliography{references}

\begin{addendum}

\item This work was supported by the GROWTH (Global Relay of Observatories Watching Transients Happen) project funded by the National Science Foundation under PIRE Grant No 1545949. GROWTH is a collaborative project among California Institute of Technology (USA), University of Maryland College Park (USA), University of Wisconsin Milwaukee (USA), Texas Tech University (USA), San Diego State University (USA), University of Washington (USA), Los Alamos National Laboratory (USA), Tokyo Institute of Technology (Japan), National Central University (Taiwan), Indian Institute of Astrophysics (India), Indian Institute of Technology Bombay (India), Weizmann Institute of Science (Israel), The Oskar Klein Centre at Stockholm University (Sweden), Humboldt University (Germany), Liverpool John Moores University (UK) and University of Sydney (Australia). 

Based on observations obtained with the Samuel Oschin Telescope 48-inch and the 60-inch Telescope at the Palomar Observatory as part of the Zwicky Transient Facility project. ZTF is supported by the National Science Foundation under Grant No. AST-1440341 and a collaboration including Caltech, IPAC, the Weizmann Institute for Science, the Oskar Klein Center at Stockholm University, the University of Maryland, the University of Washington (UW), Deutsches Elektronen-Synchrotron and Humboldt University, Los Alamos National Laboratories, the TANGO Consortium of Taiwan, the University of Wisconsin at Milwaukee, and Lawrence Berkeley National Laboratories. Operations are conducted by Caltech Optical Observatories, IPAC, and UW.  The work is partly based on the observations made with the Gran Telescopio Canarias (GTC), installed in the Spanish Observatorio del Roque de los Muchachos of the Instituto de Astrofisica de Canarias, in the island of La Palma. One of us also acknowledges all co-Is of our GTC proposal.
 
The KPED team thanks the National Science Foundation and the National Optical Astronomical Observatory for making the Kitt Peak 2.1-m telescope available. We thank the observatory staff at Kitt Peak for their efforts to assist Robo-AO KP operations. The KPED team thanks the National Science Foundation, the National Optical Astronomical Observatory, the Caltech Space Innovation Council and the Murty family for support in the building and operation of KPED. In addition, they thank the CHIMERA project for use of the Electron Multiplying CCD (EMCCD).

SED Machine is based upon work supported by the National Science Foundation under Grant No. 1106171
The ZTF forced-photometry service was funded under the Heising-Simons Foundation grant \#12540303 (PI: Graham).

M.~W.~Coughlin acknowledges support from the National Science Foundation with grant number PHY-2010970.
S.~Anand gratefully acknowledges support from the GROWTH PIRE grant (1545949).
Part of this research was carried out at the Jet Propulsion Laboratory, California Institute of Technology, under a contract with the National Aeronautics and Space Administration. 
E.C. Kool acknowledges support from the G.R.E.A.T research environment and the Wenner-Gren Foundations.
F. Foucart gratefully acknowledges support from NASA through grant 80NSSC18K0565, from the NSF through grant PHY-1806278, and from the DOE through CAREER grant DE-SC0020435.

 \item[Competing Interests] The authors declare no competing interests.
\item[Contributions] SA and MWC were the primary authors of the manuscript. MMK is the PI of GROWTH and the ZTF EM-GW program. MB, ASC, and FF led the theory and modeling. TA, MA, NG, IA, and LPS support development of the GROWTH ToO Marshal and associated program. TA, RS, JS, SBC, VZG, AKHK, HK, ECK, PM, and SR contributed to candidate scanning, vetting, and classification. EB leads the ZTF scheduler and associated interfacing with the ToO program. BB provided interpretation of the asteroid candidates. MDC, AJC, YH, RS, AFV provided GTC data and associated analysis. KD and MJH provided P200 follow-up. RGD, DAD, MF, SRK, ES and RR provided KPED data. MR and RW provided SEDM data. CF, MJG, RRL, FJM, PM, MP, PR, BR, DLS, RS, MTS, and RW are ZTF builders. All authors contributed to edits to the manuscript.

 \item[Correspondence] Correspondence and requests for materials
should be addressed to Michael Coughlin~(email: cough052@umn.edu) and Shreya Anand~(email: sanand@caltech.edu).
\end{addendum}

\clearpage
\newpage

\begin{methods}

\section{Gravitational-wave candidates}
\label{sec:events}

LIGO/Virgo S200105ae \citeNew{gcn26640}, a candidate NSBH event which occurred at 2020-01-05 16:24:26.057 UTC, was discovered by the Advanced LIGO-Livingston detector, with Virgo also observing at the time. The event was initially reported as having 97\% terrestrial probability, with a false alarm rate (FAR) of 24 per year, and therefore not generally of interest for follow-up. However, the LIGO and Virgo Collaborations reported that the significance was likely grossly underestimated as a single-instrument event, and the presence of a chirp-like structure in the spectrograms gave confidence in it being a real event \citeNew{gcn26640, gcn26657}. Unlike other NSBH events, this trigger initially had p$_{\mathrm{remnant}} > 0$\%; this parameter indicates the probability of whether there is remnant matter outside of the merger that could generate an electromagnetic transient counterpart \citeNew{ChGh2019, Foucart:2018rjc}. Similar to GW190425 \citeNew{AbEA2019}, as a single detector event, the 90\% credible region spans 7720\,deg$^2$, with an all-sky averaged distance to the source of $265 \pm 81$\,Mpc. After our observations on the three following nights were complete, a new LALInference skymap was released \citeNew{gcn26688}.  The LALInference map slightly reduced the 90\% area to 7373\,deg$^2$ (while making the 50\% area larger), modified the all-sky averaged distance to the source to $283 \pm 74$\,Mpc, and shifted more of the probability to be uniform across the lobes (including the one near the sun, which was at $\sim$ 19\,hr in RA and $\sim -22^\circ$ in declination at the time of the trigger, see \ref{fig:skymap}). Further parameter estimation maintained that the merger was likely to have contained one object with component masses $<$\,3 $M_\odot$, and therefore likely to be a neutron star ($>98$\% probability), but significantly reduced the estimated remnant probability (p$_{\mathrm{remnant}} < 1$\%).

LIGO/Virgo S200115j \citeNew{gcn26759}, a candidate NSBH event which occurred at 2020-01-15 04:23:09.742 UTC, was discovered by the two Advanced LIGO interferometers and the Advanced Virgo interferometer. This event was classified as a ``MassGap" event, with HasNS $>99$\%, indicating that one component's mass fell into the range between 3 and 5 solar masses, and the other component was $<$\,3 $M_\odot$, and therefore likely to be a neutron star, respectively. Although S200115j initially had a non-zero terrestrial probability, its revised classification reflected that the trigger was astrophysical (MassGap $>99$\%), with a FAR of 1 per 1513 years.  As a three-detector localized event, the skymap was better-constrained than for S200105ae, spanning 908\,deg$^2$ (at 90\% confidence). Additionally, it contained two disjointed lobes, one in each hemisphere, and had a median distance of $331 \pm 97$\, Mpc.  Considering all of these factors, along with the remnant probability p$_{\mathrm{remnant}}$ = 8.7\%, we chose to trigger our program for ZTF follow-up and obtained target-of-opportunity (ToO) observations. Nearly three days later, an updated LALInference skymap reduced the 90\% credible region to 765\,deg$^2$ and shifted most of the probability to the southern-most tip of the lower lobe \citeNew{gcn26807}, see \ref{fig:skymap}. The median distance was only slightly modified to $340 \pm 79$\, Mpc.  This update also distinguished S200115j from other NSBH candidates as an exceptional event for electromagnetic follow-up, with a p$_{\mathrm{remnant}} > 99$\% [ref.~\citeNew{gcn26807}].

\section{Observing Plan}
\label{sec:observing}

\subsection{S200105ae}

S200105ae was detected by LIGO and Virgo during the morning Palomar time on 2020-01-05 UT \citeNew{gcn26640}. Because it was originally identified as having a FAR above the threshold for automated public release, the skymap was not released until the following day. On 2020-01-06, beginning at 02:21:59 UT (hereafter night~$1$), only $\sim 2$\,\% of the localization was covered serendipitously by ZTF routine survey operations\citeNew{Bellm2018,Graham2018,DeSm2018,MaLa2018}, which have 30\,s observations, emphasizing that the delay in the skymap may have been a critical loss to the chances of detection for any fast fading counterparts. 

On 2020-01-07 UT (night~$2$) following the belated publication of the alert by LIGO and Virgo, we adopted a survey strategy of $g$- and $r$-band exposure blocks with 180\,s exposures for ZTF. The length of the exposures was chosen to balance both the depth required for a relatively distant event and the sky area requiring coverage; specifically, we optimize the exposure times to be as long as possible while covering the 90\% sky area consistent with the GW event observable from Palomar and in two filters within the night. We used \texttt{gwemopt} \citeNew{CoTo2018,CoAn2019}, a codebase designed to optimize telescope scheduling for GW follow-up, to schedule the observations. The schedule is designed such that fields have reference images available to facilitate image subtraction, as well as a 30 minute gap between the observations in $g$- and $r$-bands to identify and remove moving objects. These observations were submitted from the GROWTH ToO Marshal \citeNew{CoAh2019}, which we use to ingest alerts and plan observations. 

Due to poor weather conditions at Palomar, the limiting magnitudes in the first block of night~$2$ were shallower than expected at a 5\,$\sigma$ median depth of $m_\textrm{AB} =19.5$ in $g$- and $r$-bands (see \ref{fig:limmag}), and the second block originally scheduled for the same night was subsequently cancelled because of this \citeNew{gcn26662}. Combining the serendipitous and ToO observations, we covered 2200 deg$^2$, corresponding to about 44\% of the initial BAYESTAR and 35\% of the final LALInference maps on night~$2$. We adopted a similar strategy on night~$3$ (2020-01-08 UT), and improved weather led to deeper limits, with a 5\,$\sigma$ median depth of $m_\textrm{AB}=20.2$ in $g$- and $r$-bands \citeNew{gcn26673}. Combining the serendipitous and ToO observations, we covered 2100 deg$^2$ on night~$3$, corresponding to about 18\% of the initial BAYESTAR and 23\% of the LALInference maps. In total, over the 3 nights, we covered 3300\,deg$^2$, corresponding to about 52\% of the initial BAYESTAR and 48\% of the LALInference maps.

\subsection{S200115j}

The skymap for S200115j was released during Palomar nighttime on 2020-01-15 UT; we triggered ToO observations with ZTF and were on-sky within minutes. We employed the greedy-slew algorithm, same as for S200105ae, taking 300\,s exposures in $g$- and $r$-bands \citeNew{gcn26767}. Because the fields were rapidly setting by the time the skymap arrived, we were only able to cover 36\% of the skymap in our ToO observations on that night. Poor weather and seeing conditions prevented us from triggering the following night (2020-01-16 UT). The subsequently released LALInference skymap shifted the innermost probability contour to the Southern lobe \citeNew{gcn26807}, which was largely inaccessible to ZTF. While we were unable to obtain further triggered observations due to poor weather, our total serendipitous and triggered coverage within three days of the merger was 1100\,deg$^2$, corresponding to about 35\% probability of the initial BAYESTAR map and 22\% probability of the final LALInference map.

Other teams also performed synoptic follow-up of these two events~\citeNew{gcn26640,gcn26646,gcn26687,gcn26755,gcn26786,gcn26794,gcn26820}.

\section{Candidates}
\label{sec:candidates}

For a transient-event to be considered an ``alert,'' a source extracted from
 a difference image must satisfy the following criteria:
\begin{enumerate}
\item have a signal-to-noise ratio (SNR) $\geq 5$ in positive or negative flux;
\item PSF-fit magnitude $\leq 23.5$ mag;
\item number of bad pixels in 5x5 pixel region centered on transient position is $\leq 4$ pixels;
\item FWHM of source profile is $\leq 7$ pixels (where 1 pixel
 $\approx$ 1 arcsec);
\item source elongation (ratio A/B of ellipse from isophotal fit) is $\leq 1.6$;
\item the difference between flux measurements in a fixed aperture and the
 PSF-fit (${\mathrm{mag}}_{\mathrm{diff}} = {\mathrm{Aper}}_{\mathrm{mag}} - {\mathrm{PSF}}_{\mathrm{mag}}$) falls in the range:
 $-0.4\leq {\mathrm{mag}}_{\mathrm{diff}}\leq 0.75$. 
 \end{enumerate}
 For details, see Ref.~\citeNew{MaLa2018} for
 alert packet contents and Ref.~\citeNew{PaBe2018} for the ZTF alert
 distribution system.
 Hundreds of thousands of alerts are produced by ZTF every night,
 and the reader can find nightly alert collections in the ZTF alert
 archive (\url{https://ztf.uw.edu/alerts/public/}).
 
To be considered as candidates, transients must have positive residuals after image subtraction, i.e. they must have brightened relative to the reference image. We require reported transients to have at least two detections separated by at least 15 minutes to remove potential asteroids and other moving objects. In order to remove contributions from likely non-transient point sources (stars in our Galaxy and distant QSOs), we remove any candidates located less than 2\arcsec~from the Pan-STARRS1 point source catalog (PS1 PSC \citeNew{TaMi2018}), relying on star/galaxy classification as described in Ref.~\citeNew{2017AJ....153...73M}. We exclude candidates shown to be image artifacts after close inspection. We also remove any events that have detections prior to the trigger or are outside the 95\% contour in the localization. The progression in reduction of alerts to be considered for three representative nights covering the events discussed in this paper is shown in \ref{fig:table_alert}.

For cross-validation purposes, we use three forms of candidate selection, lightcurve filtering, and visualization tools: (i) the GROWTH Marshal \citeNew{Kasliwal2018}, a web-based dynamic portal for accessing transients (ii) the \texttt{Kowalski} alert archive (\url{https://github.com/dmitryduev/kowalski}) \citeNew{DuMa2019}, and (iii) the AMPEL alert archive (\url{https://github.com/AmpelProject}) \citeNew{Nordin:2019kxt,2018PASP..130g5002S}. For our realtime human vetting involving candidates from (i), we selected candidates exhibiting interesting $g$-$r$ color initially or rapid photometric evolution. Candidates retrieved via Kowalski and AMPEL (ii and iii) were all manually inspected and announced via GCN notice. As a final check, we performed a late-time Kowalski query within both event skymaps for candidates passing the above criteria, whose forced photometry lightcurves evolved faster than 0.3 mags/day, and with a baseline of $<$10 days between the first and last detection.

\section{Observation-Based NSBH Constraints}
\label{sec:constraints}

In this section, we outline a methodology for converting observational upper limits to constraints on the properties of the associated kilonova and the merging binary. Although our upper limits lack the depth required for placing meaningful constraints on the emission from both of these NSBH mergers, and we covered less than 50\% of the skymap in each case, we show that scientifically useful constraints are within reach of ZTF and similar facilities. We first illustrate how to analyze the detectability of kilonovae in a model-independent way using field-by-field ZTF pointings and a survey simulation software. Then, using a new grid of kilonova spectra tailored to NSBH mergers, we show that observations attaining a median depth of m$_\mathrm{AB}\sim$22 with improved coverage could rule out certain portions of the M$_\mathrm{ej}$--$\theta_\mathrm{obs}$ parameter space, translating to constraints on the mass ratio/NS radius/BH spin.  We describe our methodologies in detail, below. 

\subsection{Model-independent constraints}
\label{sec:model_independent}

We begin with a simple, generic model to place the observational limits in context. For this purpose, we use \texttt{simsurvey} \citeNew{FeNo2019}, a software package initially designed and used for assessing the rates of transient discovery in surveys such as ZTF by accounting for both transient and observational parameters. We adopt a toy model for transients here, injecting transients that begin at a particular absolute magnitude and decline at a certain rate measured in magnitudes per day (distributed between $-$1.0 mag/day and 1.0 mag/day, with negative decay rates corresponding to rising sources). 
We assume the transients have the same luminosity in both $g$- and $r$-band, and inject them in sky locations and distances consistent with the GW skymaps. Our results show that ZTF would be sensitive to rising or fading kilonovae brighter than M$\sim$-17.5 within the skymap of S200105ae, and fading kilonovae brighter than M$\sim$-17 within the skymap of S200115j. Losses in efficiency in general are due to our requirements that they are ``detected'' at least once within the fields we observed with ZTF; for this study, we are using both ToO and serendipitous ZTF observations from up-to 72\,hours following the merger, including time- and field-dependent limiting magnitudes from those observations.  We assume that the simulated transients evolve at the same rate during those 72\,hours. However, deeper observations of future NSBH mergers could lead to stronger statements about the minimum luminosity and maximum evolution rate of a kilonova associated with a given GW event.  In the future, as the number of NSBH merger detections increases, \texttt{simsurvey} could be used to empirically estimate the rates and luminosity function of kilonovae from NSBHs\citeNew{KaAn2020}. 

Figure~\ref{fig:efficiency} shows the percentage of transients that should be identified consistent with the LALInference skymaps for both events, parameterized by their peak absolute magnitude and decline rate.

\subsection{Ejecta mass and binary parameter constraints}
\label{sec:kilonovae}

We combine $g$- and $r$-band upper limits of S200105ae and S200115j with KN models to place constraints on the possible EM counterpart to these NSBH mergers \citeNew{MeMa2010,RoKa2011,Ro2015,KaMe2017}. We use the Monte Carlo radiative transfer code \textsc{possis} \citeNew{Bulla2019} and create a grid of spectra from which $g$- and $r$-band light curves can be extracted and compared to observations. In particular, we explore a 2D-geometry and predict light curves for eleven different viewing angles, from pole (face-on, $\cos\theta_\mathrm{obs}=1$) to equator (edge-on, $\cos\theta_\mathrm{obs}=0$).

While KN models published using \textsc{possis} have so far been focused on BNS mergers, here we present a new grid more tailored to NSBH mergers. We adopt a geometry similar to that in Figure 4 of Ref.~\citeNew{Kawaguchi2020} with two distinct ejecta components: one representing the dynamical ejecta and one the post-merger ejecta. The dynamical ejecta are characterized by a mass $M_\mathrm{ej,dyn}$, concentrated within an angle $\pm\phi$ about the equatorial plane, with velocities from 0.1 to 0.3\,c and are lanthanide-rich in composition (see Ref.~\citeNew{Bulla2019} for more details on the adopted opacities).  For simplicity, we assume a 2D geometry, where the dynamical ejecta cover an angle $2\pi$ in the azimuthal direction; we note that this is just an approximation and numerical simulations~\citeNew{kyutoku:2015,FoucartBhNs2016} suggest that this component might cover only $\sim$ half of the plane (i.e. a crescent rather than a torus). The post-merger ejecta are modelled as a spherical component with mass $M_\mathrm{ej,pm}$, extending from 0.025 to 0.1\,c and with a composition intermediate between lanthanide-poor and lanthanide-rich material \citeNew{Dietrich2020}. Below we discuss the effect of the wind composition on the derived constraints. A density profile scaling as $\rho\propto r^{-3}$ is assumed for both components. Spectra for this new grid are made available at \url{https://github.com/mbulla/kilonova_models}. 

To place constraints on the ejected material, we fix $\phi=30^\circ$ and run a grid of 81 models with varying ejecta masses for the two components: $M_\mathrm{ej,dyn}, M_\mathrm{ej,pm} \in [0.01,0.09]\,M_\odot$ (step size $0.01\,M_\odot$). The simulated light curves show a strong dependence on the viewing angle, with increasingly fainter KNe when moving the observer from the pole ($\cos\theta_\mathrm{obs}=1$) to the equator ($\cos\theta_\mathrm{obs}=0$). In particular, orientations in the equatorial plane are on average $2-3$\,mag fainter in $g$-band than those along the polar direction due to the blocking effect of the dynamical ejecta \citeNew{Bulla2019,Kawaguchi2020b}. This blocking effect may be in part a consequence of the choice of an axisymmetric outflow geometry. For a more realistic geometry of the dynamical ejecta, the post-merger ejecta would remain unobscured for some equatorial observers. 3D radiation transfer simulations with a non-axisymmetric dynamical ejecta may thus provide stronger constraints on the ejected mass for at least some equatorial observers than the 2D simulations performed here. We note that the discrepancy mentioned in Ref.~\citeNew{Kawaguchi2020b} between their light curves and those in Ref.~\citeNew{Bulla2019} is now negligible following an update of \textsc{possis} where the temperature is no longer parameterized and uniform but rather calculated at each time and in each zone from the mean intensity of the radiation field. In addition, here we adopt thermalization efficiencies $\epsilon_\mathrm{th}$ from Ref.~\citeNew{Barnes2016} rather than assuming $\epsilon_\mathrm{th}=0.5$ as in Ref.~\citeNew{Bulla2019}. For instance, we obtain a $g$-band absolute magnitude of $-$15.3 mag at 1\,day for the model with $M_\mathrm{ej,dyn}=M_\mathrm{ej,pm}=0.02\,M_\odot$ viewed face-on (cf. with Figure 16 of Ref.~\citeNew{Kawaguchi2020}). \ref{fig:nsbh_lc} provides an example set of light curves in the passbands utilized in observations in this paper. The significantly brighter emission in $i$- and $z$-band compared to $g$- and $r$-band implies that better overall constraints on the kilonova emission are expected. To perform this check systematically, we present \ref{fig:iz_vs_gr}, which demonstrates the difference in peak magnitudes between $g$- and $r$-bands and $i$- and $z$-bands for the models in the NSBH grid used here. The result of brighter emission in $i$- and $z$-band compared to $g$- and $r$-band holds true across the parameter space, with peak $z$-band observations generally exceeding $g$-band by 1\,mag or more.

To demonstrate possible constraints from deeper observations, which would have been achievable under better weather conditions, we also examine constraints given by the most limiting individual pointings in each set of observations. The aim of this analysis is to guide future follow-up comparisons, showing what constraints could have been achieved should all the observations have been taken with the same depth as in the deepest field. Compared to the median values used above, individual observations reach deeper magnitudes (see open triangles in the left and middle panels of Figure~\ref{fig:kn_lc}). Results of this analysis are shown in \ref{fig:kne}, where we highlight the deepest limits for each set of observations.

The left column in \ref{fig:kne} summarizes results for S200105ae. 
The top panels show $g$- and $r$-band light curves that would be ruled out if our median limits had reached the depth of our deepest observations on each night, for different distance assumptions (209, 283 and 357 Mpc from light to dark blue). We could rule out more models at closer compared to farther distances. In particular, all the models can be ruled out by the $r$-band upper limit at $\sim$~3 days, $m_r>22.35$ mag, with no improvement found when adding the other observations. The bottom panels show what regions of the $M_\mathrm{ej,dyn}-M_\mathrm{ej,pm}$ parameter space are ruled out by observations for three different viewing angle ranges: $0.9<\cos\theta_\mathrm{obs}<1$ ($0<\theta_\mathrm{obs}<26^\circ$), $0.6<\cos\theta_\mathrm{obs}<0.7$ ($46^\circ<\theta_\mathrm{obs}<53^\circ$) and $0<\cos\theta_\mathrm{obs}<0.1$ ($84^\circ<\theta_\mathrm{obs}<90^\circ$). As expected, polar orientations are more constraining than the other ranges. In particular, our deepest observations could constrain the ejecta masses to $M_\mathrm{ej,dyn}\leq0.02\,M_\odot$ and $M_\mathrm{ej,pm}\leq0.04\,M_\odot$ for polar directions at 283 Mpc. Weaker constraints are found for orientations away from the pole, with all KNe being sufficiently faint and thus not ruled out by upper limits for an equatorial observer (bottom-left panel).

The middle column in \ref{fig:kne} shows the same analysis for S200115j. 
For S200115j, the larger distance and shallower limits lead to fewer models ruled out and thus poorer constraints in the $M_\mathrm{ej,dyn}-M_\mathrm{ej,pm}$ parameter space.
Specifically, models are ruled out only in the optimistic case of 261 Mpc and viewing angle close to the pole. For S200115j, the most (and only) constraining observations are the limits at $\sim1$~day. 

We also provide updated results for S190814bv using our NSBH-specific KN model. For S190814bv, stronger constraints can be derived even for median observing depths. These constraints are also more reliable, as observations\citeNew{Andreoni2020} covered 98\% of the LVC skymap. On the other hand, constraints on the parameter space of the binary are unlikely to provide information distinct from that extracted from GW observations, as the LVC already indicates that this event has $0\%$ probability of being EM-bright. We find that all of our KN models are ruled out for polar orientations at $\leq 267\,{\rm Mpc}$, effectively limiting the dynamical and post-merger ejecta masses to $\leq 0.01M_\odot$. This would lead to constraints on the binary parameters shown on \ref{fig:pe2}. For higher inclinations ($46^\circ \leq \theta\leq 53^\circ$), the constraints are similar to what we just obtained for deep observations of S200105ae, with limits on the binary parameters accordingly close to those displayed on \ref{fig:pe}.

\section{Data Availability}
The data that support the plots within this paper and other findings of this study are available from the corresponding author upon reasonable request.

\section{Code Availability}

Upon request, the corresponding author will provide code (primarily in python) used to produce the figures.

\end{methods}

\begin{extended_data}

\renewcommand{\thefigure}{Extended Data Figure \arabic{figure}}
\renewcommand{\figurename}{}
\setcounter{figure}{0}



\begin{figure*}[t]
 \begin{center}
 \includegraphics[width=0.97\textwidth]{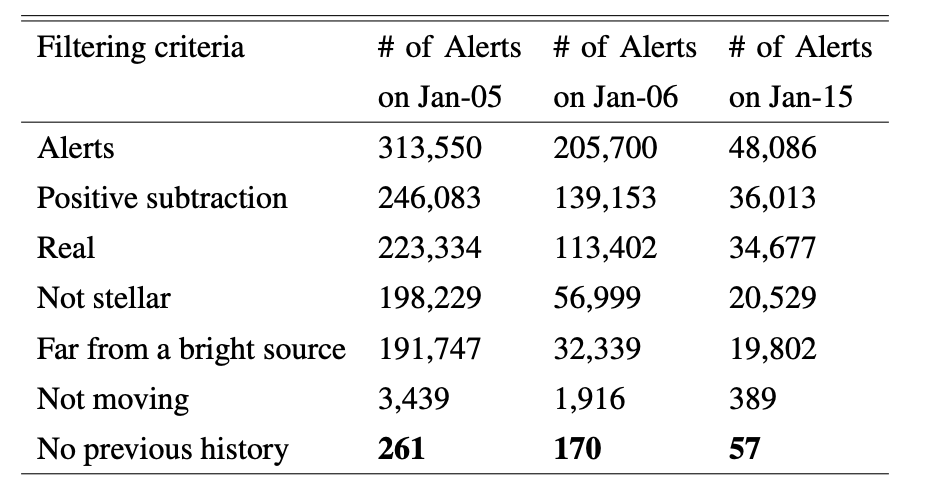}
  \caption{\textbf{Automatic preliminary filtering criteria for transient detection.} Here we show results for each step of the ZTF filtering scheme for three representative nights covering the events discussed in this paper. Each cell shows the number of candidates that successfully pass a particular filter. The number shown is the result of running a filtering step on the alerts that met previous requirements. We define as ``Real'' any alert with a real-bogus score greater than 0.25 and ``not moving'' the candidates that have more than two detections separated by at least 15 minutes. The highlighted numbers represent the amount of candidates that required further vetting, as described in Section \ref{sec:candidates}.}
 \label{fig:table_alert}
 \end{center}
\end{figure*}

\begin{figure*}[t]
  \includegraphics[width=1\textwidth]{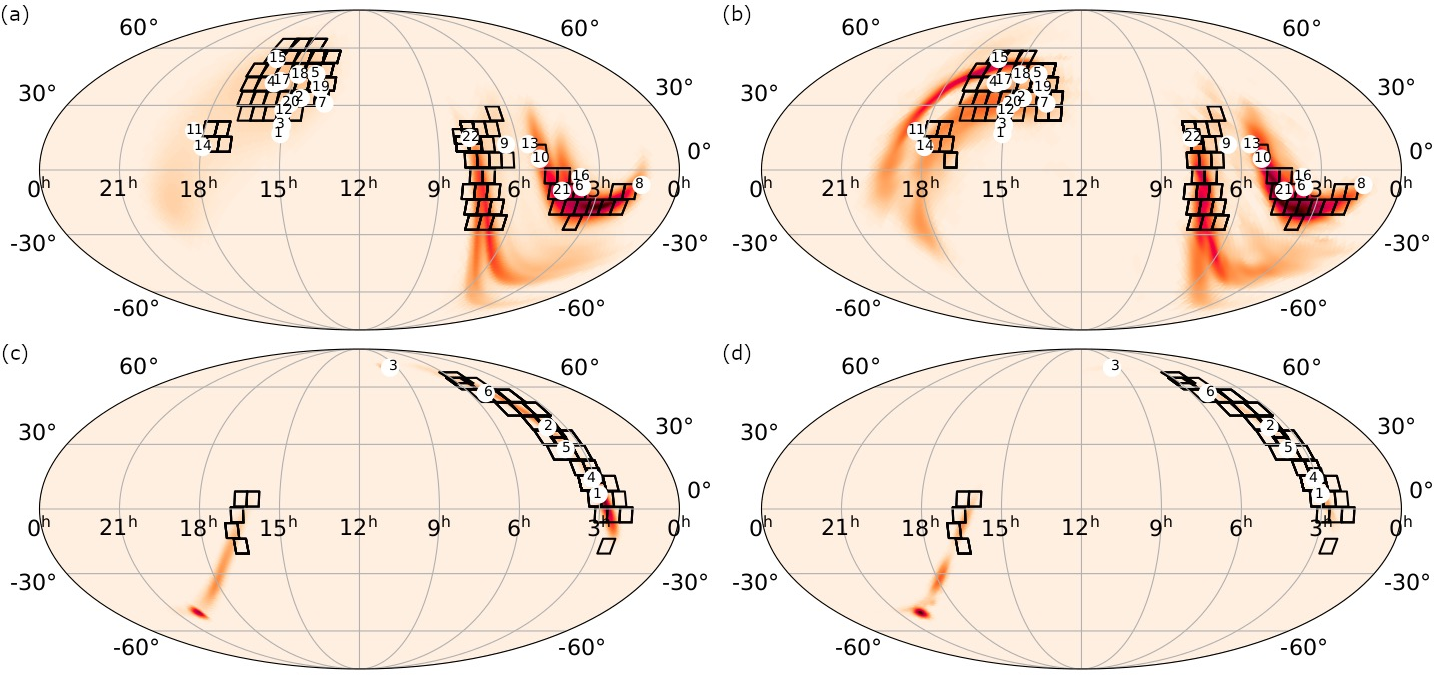}
  \caption{\textbf{ZTF coverage and candidates discovered within skymap.} Top row: Coverage of S200105ae, showing the tiles on the 90\% probability region of the initial BAYESTAR (a) and final LALInference (b) skymaps. The color intensity is proportional to the 2-D probability. The mapping of candidates to numbers is 1: ZTF20aaervoa, 2: ZTF20aaertpj, 3: ZTF20aaervyn, 4: ZTF20aaerqbx, 5: ZTF20aaerxsd, 6: ZTF20aafduvt, 7: ZTF20aaevbzl, 8: ZTF20aaflndh, 9: ZTF20aaexpwt, 10: ZTF20aafaoki, 11: ZTF20aafukgx, 12: ZTF20aagijez, 13: ZTF20aafanxk, 14: ZTF20aafujqk, 15: ZTF20aagiiik, 16: ZTF20aafdxkf, 17: ZTF20aagiipi, 18: ZTF20aagjemb, 19: ZTF20aafksha, 20: ZTF20aaertil, 21: ZTF20aafexle and 22: ZTF20aafefxe. Bottom row: Same for S200115j, with the BAYESTAR coverage shown in (c) and LALInference coverage shown in (d). The mapping of candidates to numbers is 1: ZTF20aagjqxg, 2: ZTF20aafqvyc, 3: ZTF20aahenrt, 4: ZTF20aafqpum, 5: ZTF20aafqulk, and 6: ZTF20aahakkp. We note that we include candidates up to and including within the 95\% probability region, and therefore some are outside of the fields we plot here.
  }
 \label{fig:skymap}
\end{figure*} 

\begin{figure*}[t]
 \centering
  \includegraphics[width=5.0in]{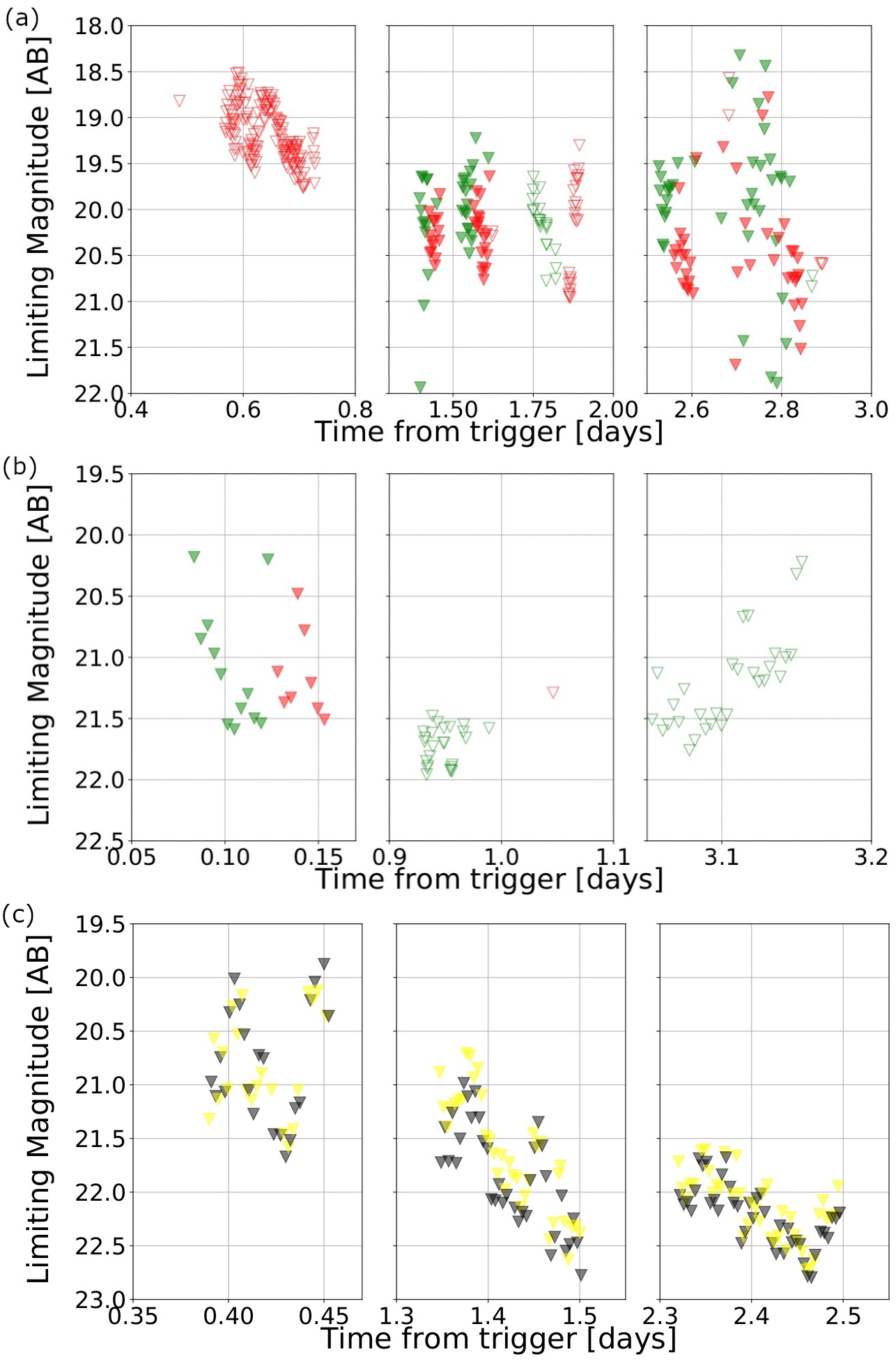}

  \caption{\textbf{Limiting magnitudes at each epoch of observations.} 5-$\sigma$ limiting magnitudes as a function of time for (a) S200105ae (ZTF), (b) S200115j (ZTF), and (c) S190814bv (DECam) with the left, middle, and right panels corresponding to observations on the first, second, and third nights for S200105ae and S190814bv and first, second, and fourth nights for S200115j. The red and green triangles correspond to the $r$- and $g$-band limits for ZTF, while the yellow and black triangles correspond to the $i$- and $z$-band limits for DECam; the open triangles correspond to serendipitous observations and closed ToO observations. The large differences in limiting magnitude from observation to observation are due to poor weather.}
 \label{fig:limmag}
\end{figure*} 

\begin{figure*}[t]
 \begin{center}
 \includegraphics[width=0.88\textwidth]{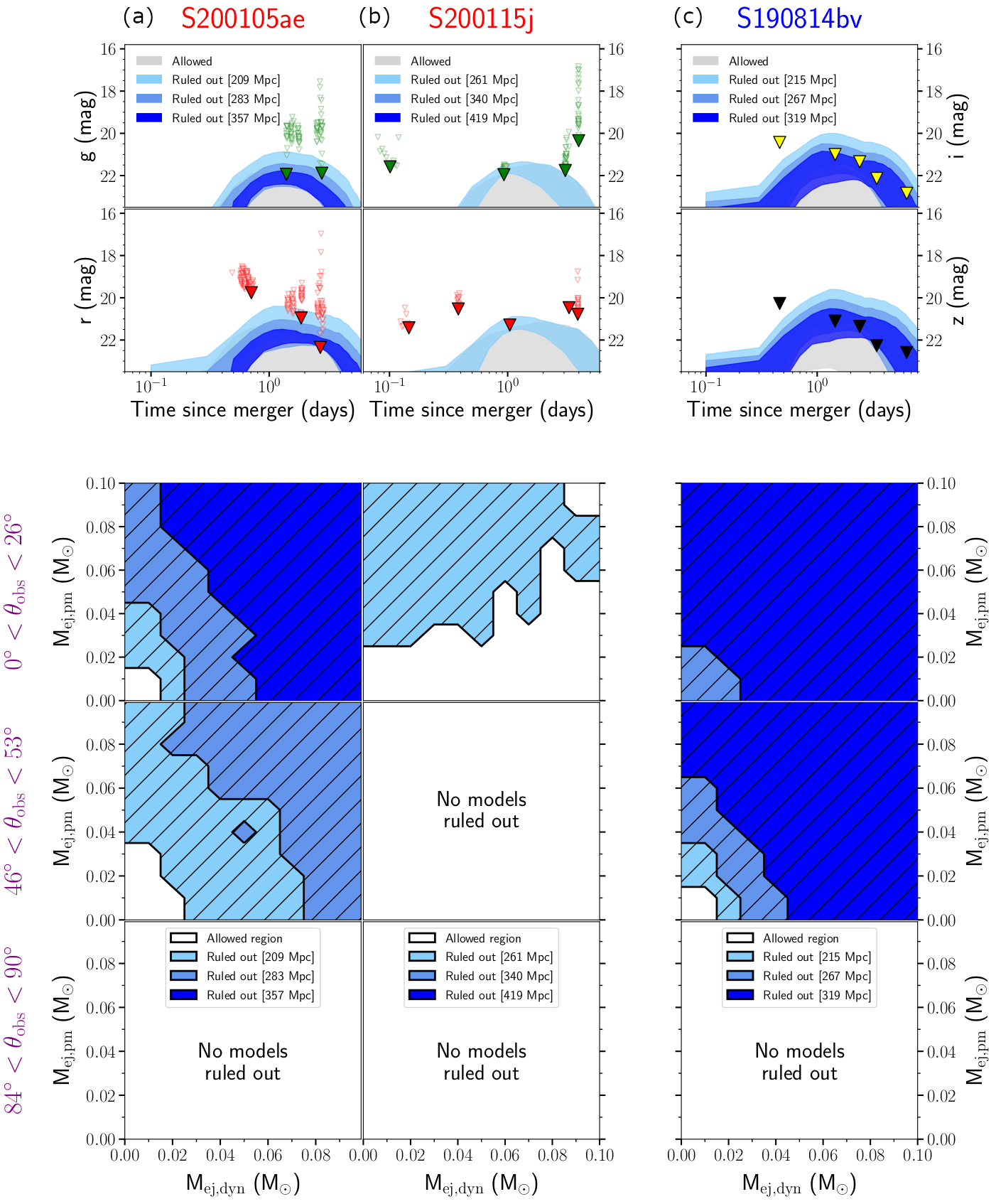}
  \caption{\textbf{Potential constraints on kilonova model parameters based on the \textit{deepest} limiting magnitudes.} We display constraints on (a) S200105ae (ZTF), (b) S200115j (ZTF) and (c) S190814bv (DECam) for the models in the NSBH grid used here. \textit{Top panels}: same as Figure~\ref{fig:kn_lc} but using the deepest (filled triangles) rather than the median limits for each set of S200105ae and S200115j observations. The panel for S190814bv is the same as in Figure~\ref{fig:kn_lc}, with all limits corresponding to the median magnitudes. \textit{Bottom panels}: regions of the $M_\mathrm{ej,dyn}-M_\mathrm{ej,pm}$ parameter space that are ruled out at different distances and for different viewing angle ranges (moving from pole to equator from top to bottom panel).}
 \label{fig:kne}
 \end{center}
\end{figure*} 

\begin{figure*}[t]
 \begin{center}
 \includegraphics[width=0.97\textwidth]{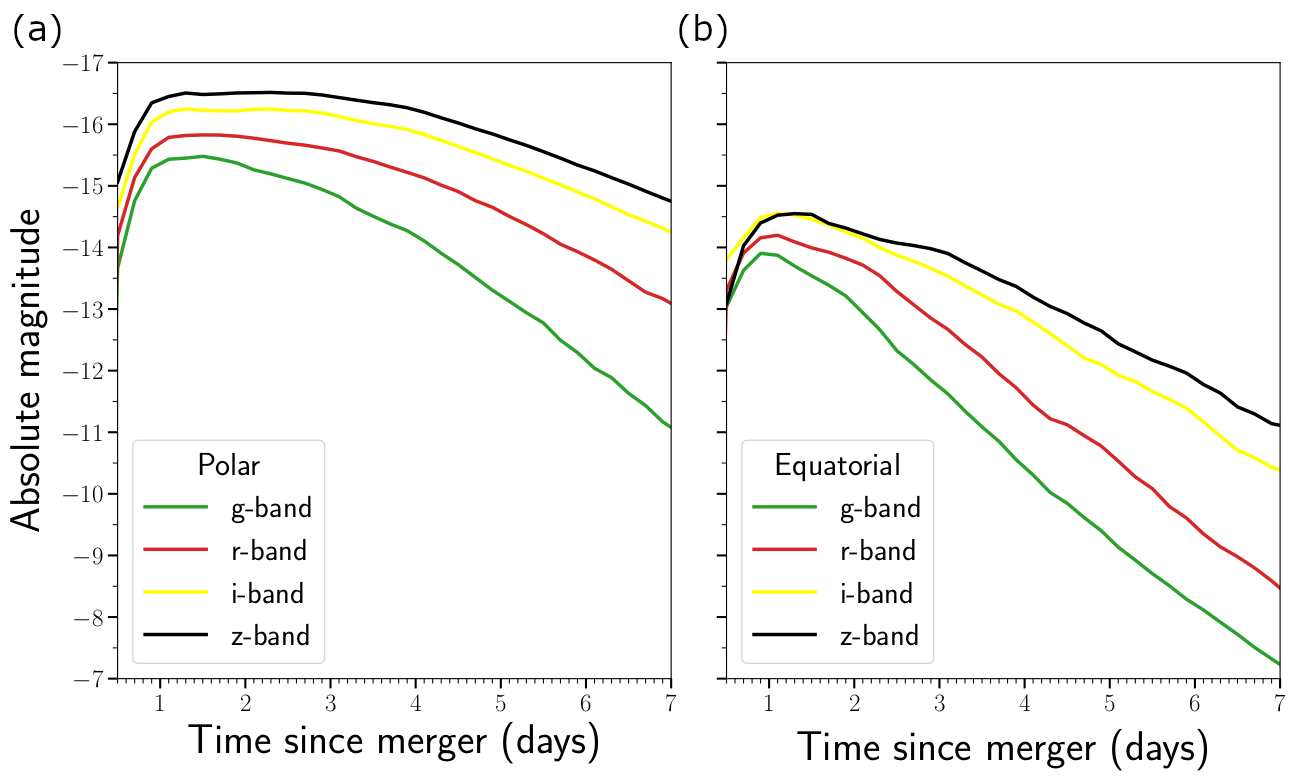}
  \caption{\textbf{Broadband NSBH lightcurve models from \textsc{possis}.} Light curves predicted with \textsc{possis} \protect\citeNew{Bulla2019} for a NSBH model with $M_{\rm dyn}=0.05M_\odot$ and $M_{\rm pm}=0.05M_\odot$ as seen from a polar (a) and equatorial (b) viewing angle.}
 \label{fig:nsbh_lc}
 \end{center}
\end{figure*}

\begin{figure*}[t]
 \begin{center}
 \includegraphics[width=0.97\textwidth]{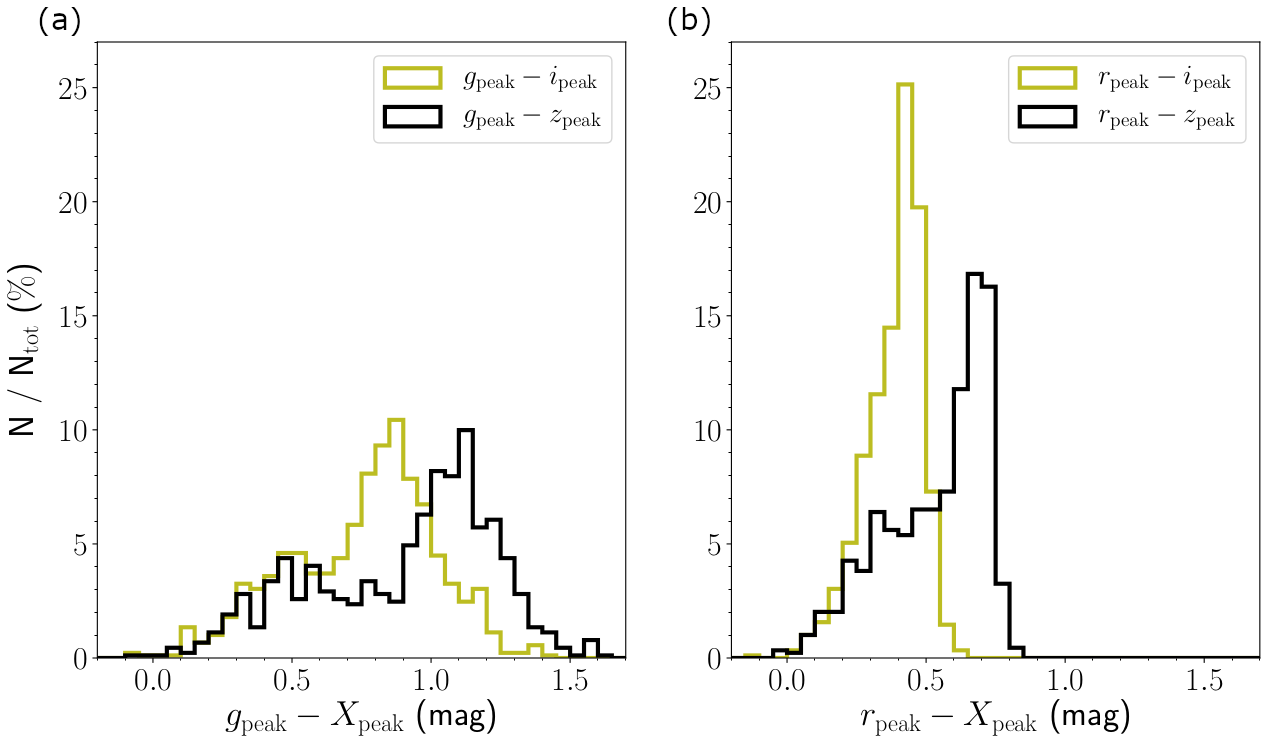}
  \caption{\textbf{Comparison of peak magnitudes between optical and near-IR bands for NSBH models.} We plot the difference in peak magnitudes between the (a) $g$-band and the near-IR $i$- and $z$-bands for the models in the NSBH grid used here. Similarly, in (b) we show the difference between $r$-band and the same near-IR bands.}
 \label{fig:iz_vs_gr}
 \end{center}
\end{figure*}

\begin{figure*}[t]
 \begin{center}
 \includegraphics[width=0.97\textwidth]{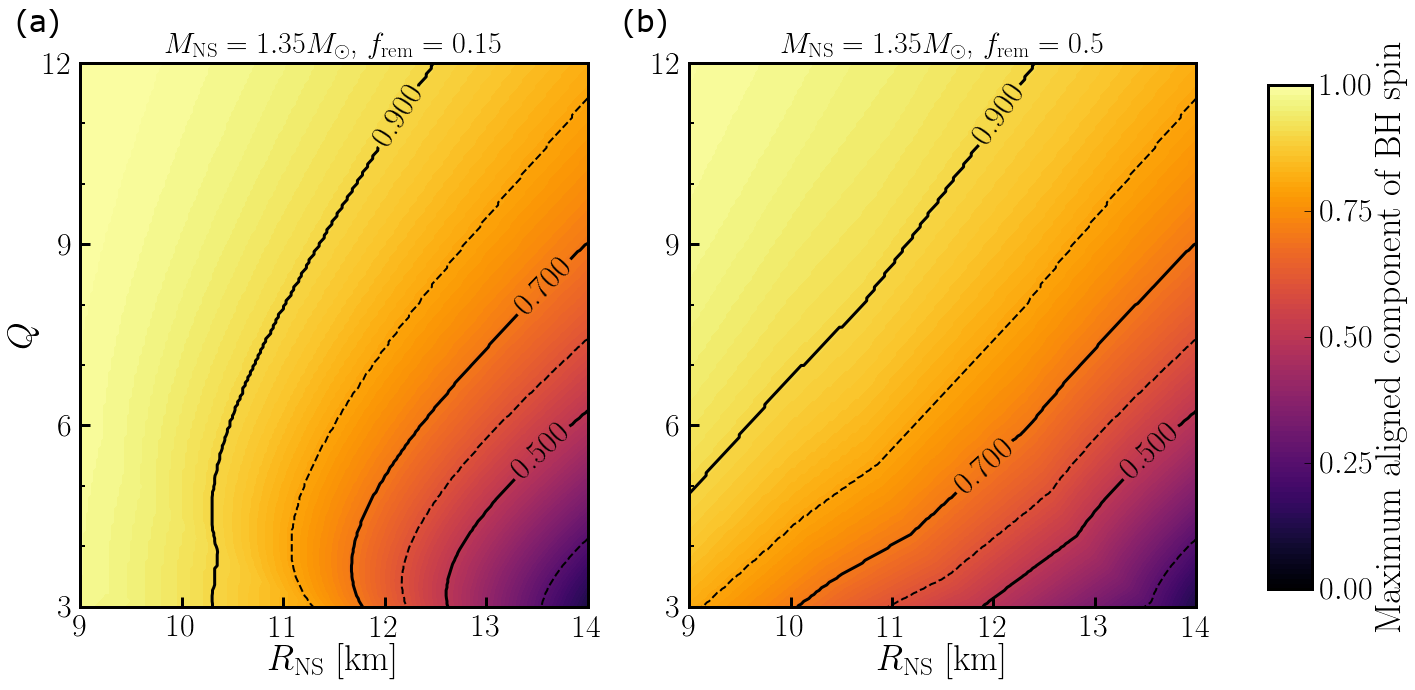}
  \caption{\textbf{Potential constraints on the parameters of a NSBH binary associated with S200105ae.} Here we assume that $M_{\rm ej,dyn}\leq 0.02M_\odot$ and $M_{\rm ej,pm}\leq 0.04M_\odot$, appropriate for the deepest observations of S200105ae in a face-on orientation. We show the maximum value of the aligned component of the BH spin as a function of the neutron star radius $R_{\rm NS}$ and the binary mass ratio $Q=M_{\rm BH}/M_{\rm NS}$. The two panels show results assuming that low (a) and high (b) fractions of the post-merger accretion disk are ejected (see text). Both plots assume $M_{\rm NS}=1.35$. Results for different neutron star masses can be estimated from this plot simply by considering a binary with the same $Q,\chi$ and compaction $M_{\rm NS}/R_{\rm NS}$.}
 \label{fig:pe}
 \end{center}
\end{figure*} 

\begin{figure}[t]
 \begin{center}
 \includegraphics[width=0.97\textwidth]{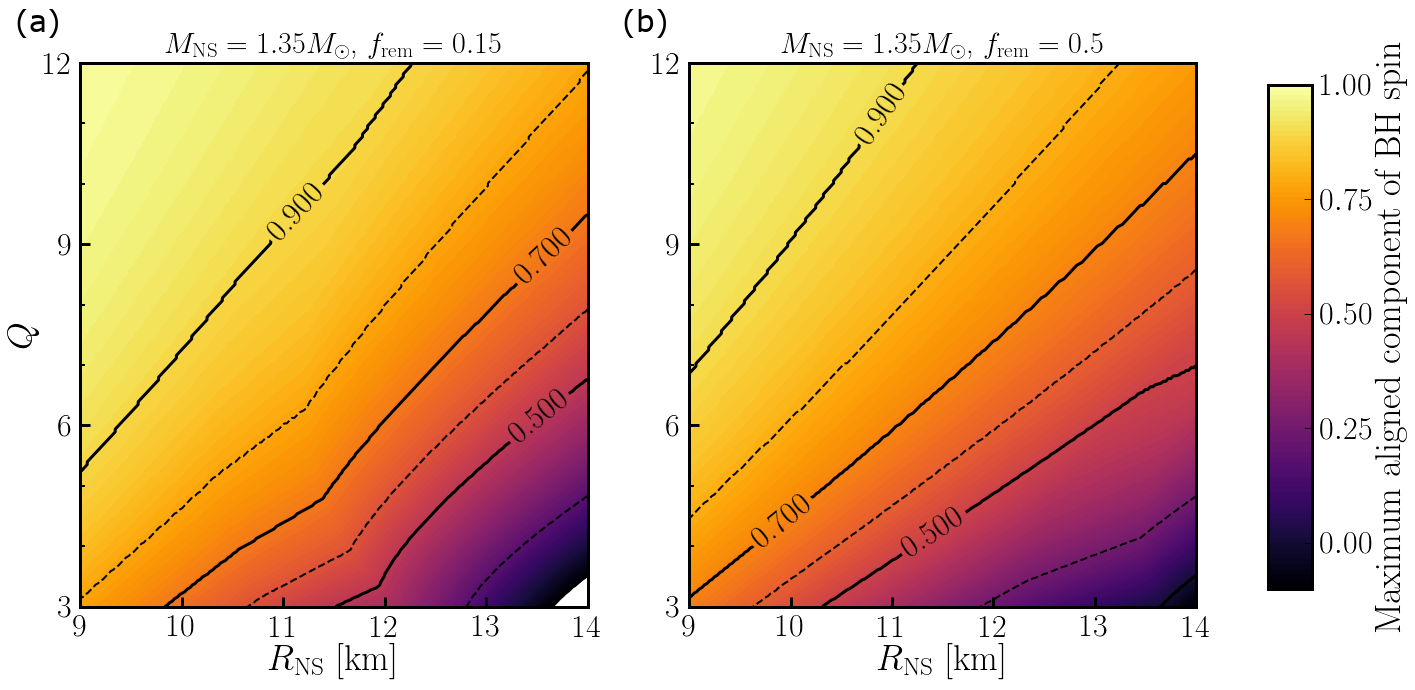}
  \caption{\textbf{Potential constraints on the parameters of a NSBH binary associated with S190814bv.} Here we assume that $M_{\rm ej,dyn}\leq 0.01M_\odot$ and $M_{\rm ej,pm}\leq 0.01M_\odot$, as appropriate for S190814bv in a face-on orientation in a similar fashion to \ref{fig:pe}, with low (a) and high (b) fractions of disk ejecta.}
 \label{fig:pe2}
 \end{center}
\end{figure} 

\begin{figure}[t]
 \begin{center}
 \includegraphics[width=0.97\textwidth]{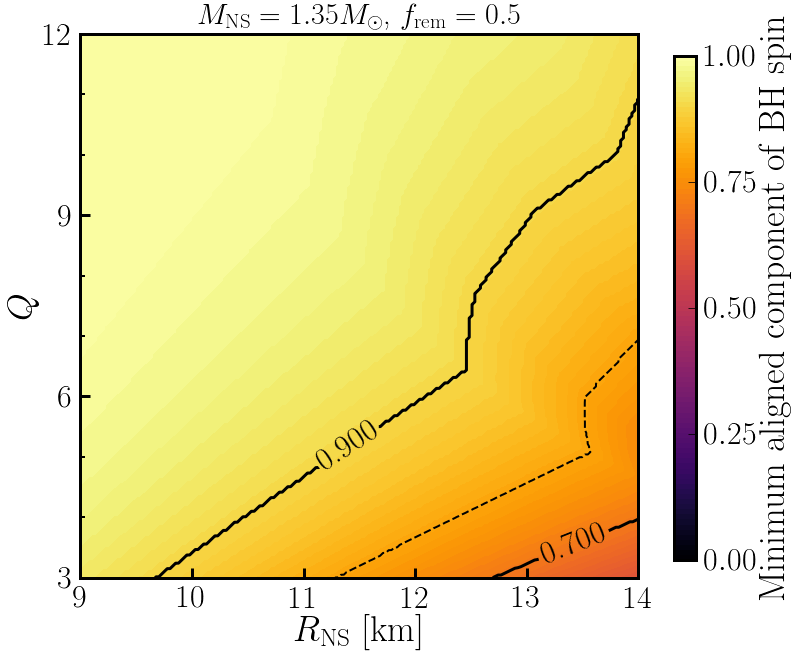}
  \caption{\textbf{Minimum aligned component of the BH spin above which we cannot rule out the presence of a kilonova.} We cannot exclude this region of parameter space because either the resulting kilonova evolves too slowly, or the ejected mass is outside of the grid of models used in this study. In this plot, we consider the worse-case scenario of $f_{\rm rem}=0.5$.}
 \label{fig:pelim}
 \end{center}
\end{figure} 

\end{extended_data}

\clearpage

\begin{supplement}

\renewcommand{\thefigure}{Supplementary Information Figure~\arabic{figure}}
\renewcommand{\figurename}{}
\setcounter{figure}{0}

\section{Observational details}
\subsection{Photometric Observations}
The ZTF observations used to discover potential candidates were primarily obtained with ToO program time, however the public survey \citeNew{Bellm2018} provided us with data as well. The nominal exposure time for the ZTF public survey is 30s while for the ToO program varies from 120-300\,s depending on the available time and sky area requiring coverage. Our first source of photometry comes from the ZTF alert production pipeline \citeNew{MaLa2018}, however for the purposes of this paper we have performed forced photometry using the package \texttt{ForcePhot}\citeNew{YaMi2019} on the candidates and reported these values. 

For S200105ae, we split the schedule into two blocks of right ascension due to the significantly displaced lobes in the skymap (see \ref{fig:skymap}), with observations lasting three hours per block. We additionally utilized the ``filter balancing" feature \citeNew{AlCo2020}, which optimizes for the number of fields that have observations scheduled in all requested filters, and employed the greedy-slew algorithm \citeNew{RaAn2019} for conducting our search. 
The ability to split the skymap in right ascension and the use of filter balancing was novel for these observations, and served to help address the previous difficulty with multi-lobed skymaps to make it possible to observe all filters requested for the scheduled fields. Previously, maps of this type created conflicts between the rising/setting times of the lobes, as well as the separation in time between each of the epochs. This problem impacts the transient filtering process as well, for example, resulting in a number of transients failing to satisfy the criteria of 15\,minutes between consecutive detections to reject asteroids. With the implementation of these features, both $g$- and $r$-band epochs were successfully scheduled for almost all fields.

For photometric follow-up we used the Gemini Multi-Object Spectrograph (GMOS-N)\citeNew{HoJo2004} on the Gemini-North 8-meter telescope on Mauna Kea, the Spectral Energy Distribution Machine (SEDM) on the Palomar 60-inch telescope \citeNew{BlNe2018}, the Wide-field Infrared Camera (WIRC)\citeNew{Wilson2003} on the Palomar 200-inch telescope, as well as telescopes that are part of the Las Cumbres Observatory (LCO) network and the Kitt Peak EMCCD Demonstrator (KPED)\citeNew{Coughlin2018}.

The LCO observations were scheduled using the LCO Observation Portal (\url{https://observe.lco.global/}), an online platform designed to coordinate observations. Our imaging plans changed case by case, however our standard requests involved 3 sets of 300s in $g$- and $r$- band with the 1-m telescopes. For fainter sources we requested 300s of $g$- and $r$- band with the 2-m telescopes. The reduced images available from the Observation Portal were later stacked and sources were extracted with the SourceExtractor package\citeNew{BeAr1996}. We calibrated magnitudes against Pan-STARRS1\citeNew{ChMa2016} sources in the field. For transients separated $<$\,8\arcsec~from their hosts, we aligned a cutout of the transient with a Pan-STARRS1 template using SCAMP\citeNew{Ber2006} and performed image subtraction with the High Order Transform of Psf ANd Template Subtraction (HOTPANTS) code \citeNew{hotpants}, an enhanced version of the method derived by Ref.~\citeNew{Ala2000}. Photometry for these candidates comes from an analogous analysis on the residual images. Furthermore, images obtained with the Liverpool telescope (LT)\citeNew{StRe2004} were reduced, calibrated and analysed in a similar fashion. 

For KPED data, our standard procedure is to stack an hour of $r$-band data and reduce the stacked images following to standard bias and flat field calibrations. The photometry is obtained following the same methods as for the LCO data. 

The photometric data obtained with GMOS-N was split in four 200\,s $g$-band images later combined and reduced with DRAGONS (\url{https://dragons.readthedocs.io/en/stable/}), a Python-base data reduction platform provided by the Gemini Observatory. The data were later calibrated using the methods described for LCO. 

Additionally, we scheduled photometric observations with the SEDM automatically through the GROWTH marshal. We acquired $g$-, $r$-, and $i$- band imaging with the Rainbow Camera on SEDM in 300s exposures. SEDM employs a python-based pipeline that performs standard photometric reduction techniques and uses an adaptation of \texttt{FPipe} (Fremling Automated Pipeline; described in detail in Ref.~\citeNew{FrSo2016}) for difference imaging. Data are automatically uploaded to the GROWTH marshal after having been reduced and calibrated.

The near-infrared data obtained with WIRC were reduced using a custom data reduction pipeline described in Ref.~\citeNew{DeHa2020}, and involved dark subtraction followed by flat-fielding using sky-flats. The images were then stacked using Swarp \citeNew{Bertin2002} and photometric calibration was performed against the 2MASS point source catalog \citeNew{Skrutsie2006}. Reported magnitudes were derived by performing aperture photometry at the location of the transient using an aperture matched to the seeing at the time of observation, including an aperture correction to infinite radius.

The photometry presented in the light-curves on this paper was corrected for galactic extinction using dust maps from Ref.~\citeNew{ScFi2011}. 

\subsection{Spectroscopic Observations}
For the candidate dataset described in Sec.~\ref{sec:candidates}, we obtained spectroscopic data using the Gran Telescopio Canarias (GTC) and Palomar observatory.
We obtained optical spectra of one set of candidates with the 10.4-meter GTC telescope (equipped with OSIRIS). Observations made use of the R1000B and R500R grisms, using typically a slit of width 1.2\arcsec. Data reduction was performed using standard routines from the Image Reduction and Analysis Facility (IRAF).

For the second set of candidates, we acquired most of our spectra with the Integral Field Unit (IFU) on SEDM, a robotic spectrograph on the Palomar 60-inch telescope \citeNew{BlNe2018}. We scheduled spectroscopic observations for our brighter (m$_{AB} < 19$) and higher priority targets using a tool on the GROWTH Marshal that directly adds the target to the SEDM queue. For each science target, the SEDM robot obtains an acquisition image, solves the astrometry and then sets the target at the center of the integral field unit field of view. At the end of exposure, the automated pysedm pipeline is run \citeNew{RiNe2019}. It first extracts the IFU spaxel tracers into a x,y,$\lambda$ cube accounting for instrument flexures; the target spectrum is then extracted from the cube using a 3D PSF model which accounts for atmospheric differential refractions. The spectrum is finally flux calibrated using the most recent standard star observation of the night, with the telluric absorption lines scaled for the target's airmass. See Ref.~\citeNew{RiNe2019} for more details on the reduction pipeline. The final extracted spectra are then uploaded to the marshal; we use the SNID software \citeNew{BlTo2007} to classify our transients. 

Using the Double Spectrograph (DBSP) on the Palomar 200-inch telescope we obtained one transient and one host galaxy spectrum during our classical observing run on 2020-01-18 UT. For the setup configuration, we use 1.0\arcsec and 1.5\arcsec slitmasks, a D55 dichroic, a B grating of 600/4000 and R grating of 316/7500.  Data were reduced using a custom PyRAF DBSP reduction pipeline (https://github.com/ebellm/pyraf-dbsp)~\citeNew{BeSe2016}. 

\section{Candidates}

\subsection{S200105ae candidates}\label{section:05aecandidates}


In this subsection, we provide brief descriptions of candidates identified within the skymap of S200105ae. Due to the poor seeing conditions and moon brightness, there were no candidates that passed all of the criteria after the second night of observations. After the third night of observations of S200105ae, we identified 5 candidates within the skymap \citeNew{gcn26673}, shown in Supplementary Information Table~\ref{table:S200105ae_spec_followup} and on \ref{fig:skymap}. In addition, we later identified and reported other candidate counterparts \citeNew{gcn26810}. A late-time query ($>1$ month after the mergers) yielded two further candidates of interest, ZTF20aafsnux and ZTF20aaegqfp, that were not already reported via Gamma-ray burst Coordinates Network (GCN). 


\begin{figure*}
    \centering
    \includegraphics[width=2.25in]{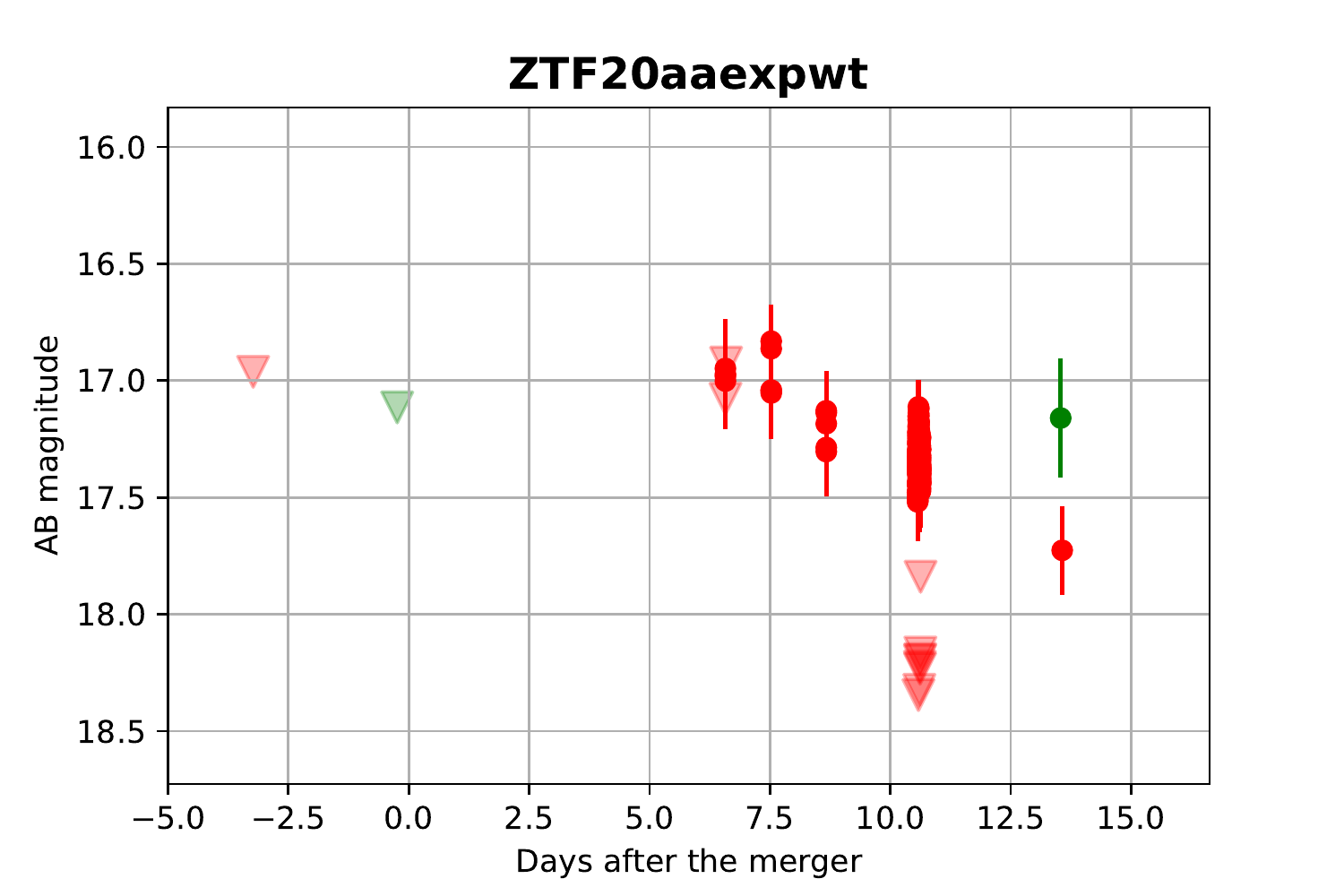}\includegraphics[width=2.25in]{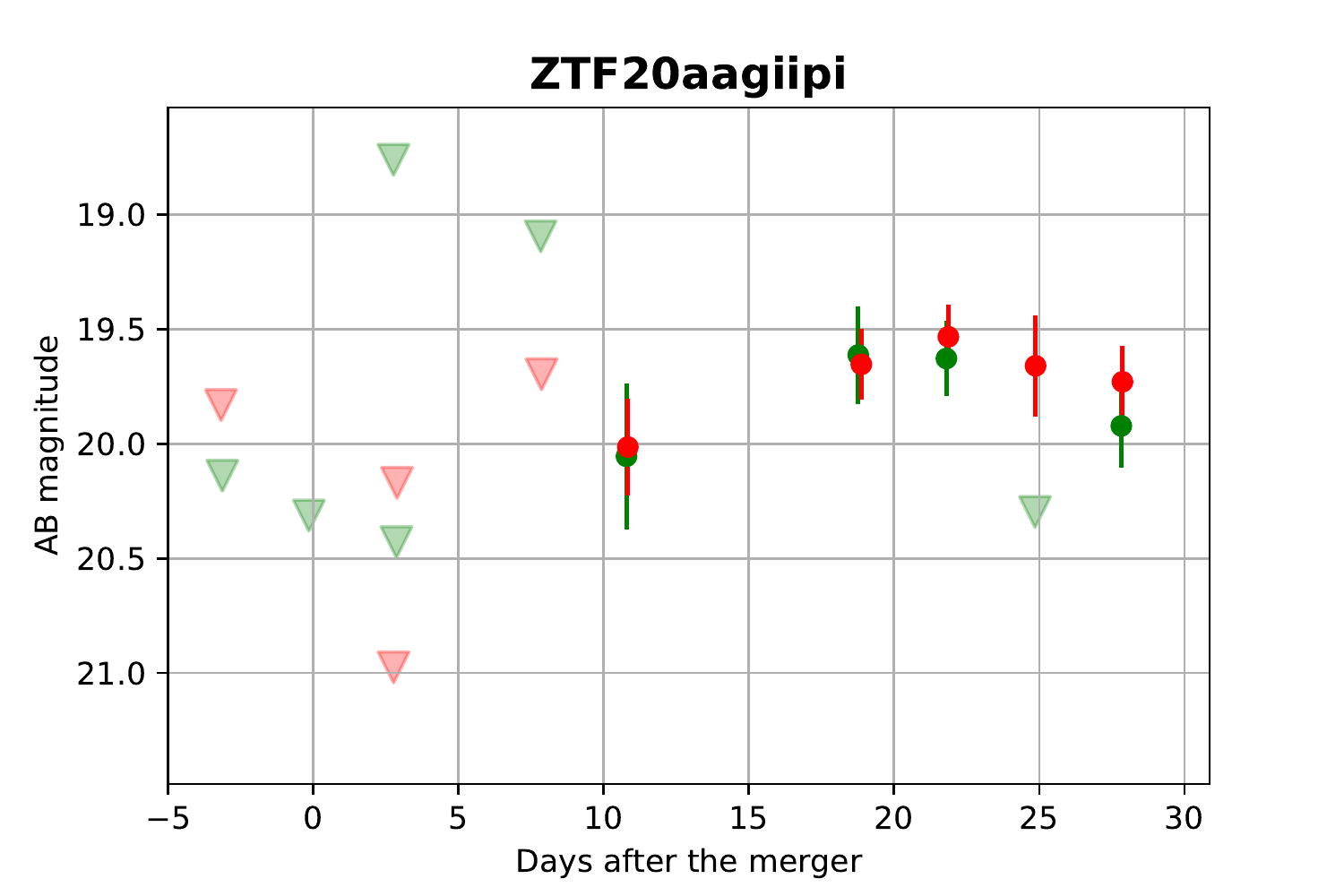}\includegraphics[width=2.25in]{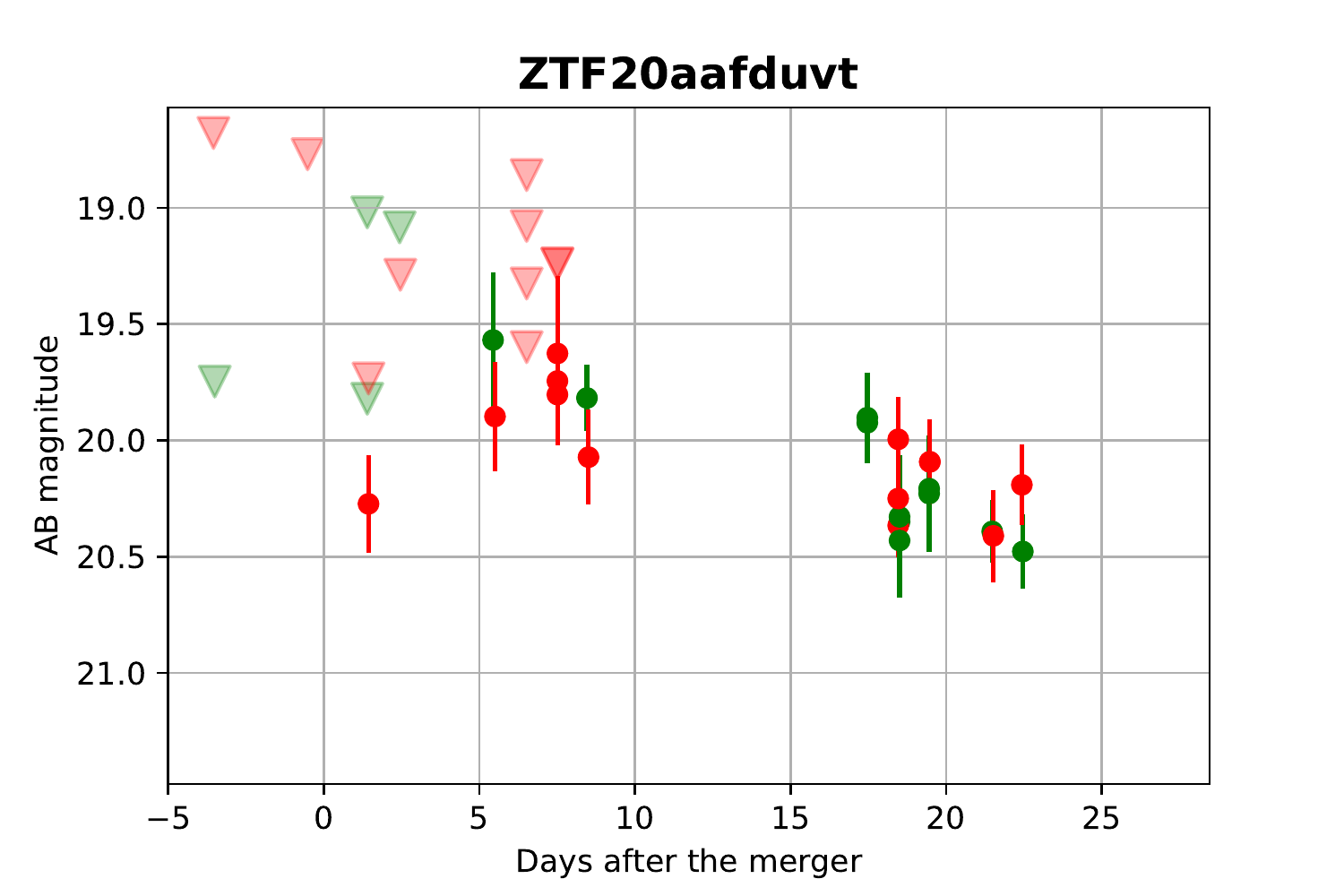}
    \includegraphics[width=2.25in]{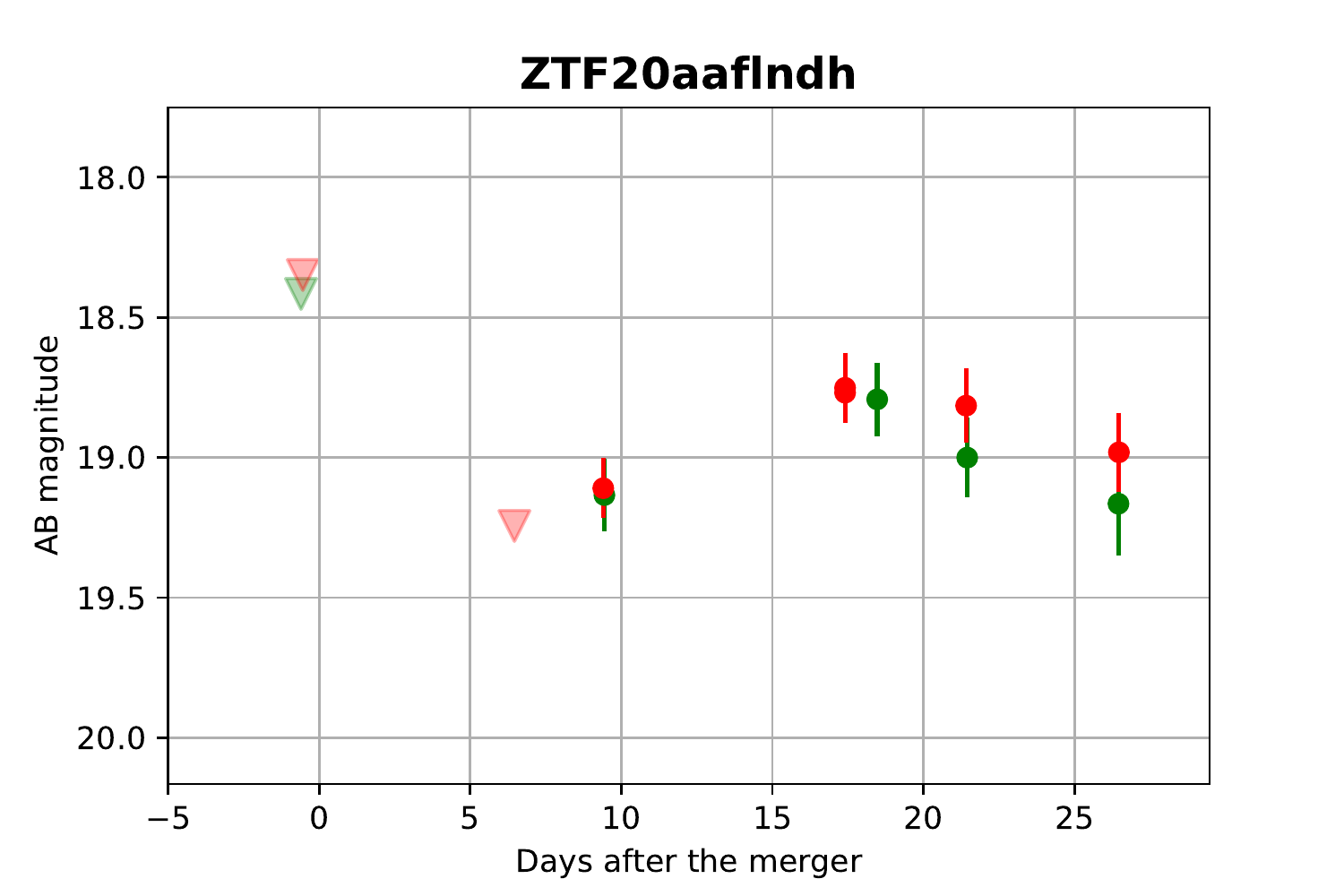}\includegraphics[width=2.25in]{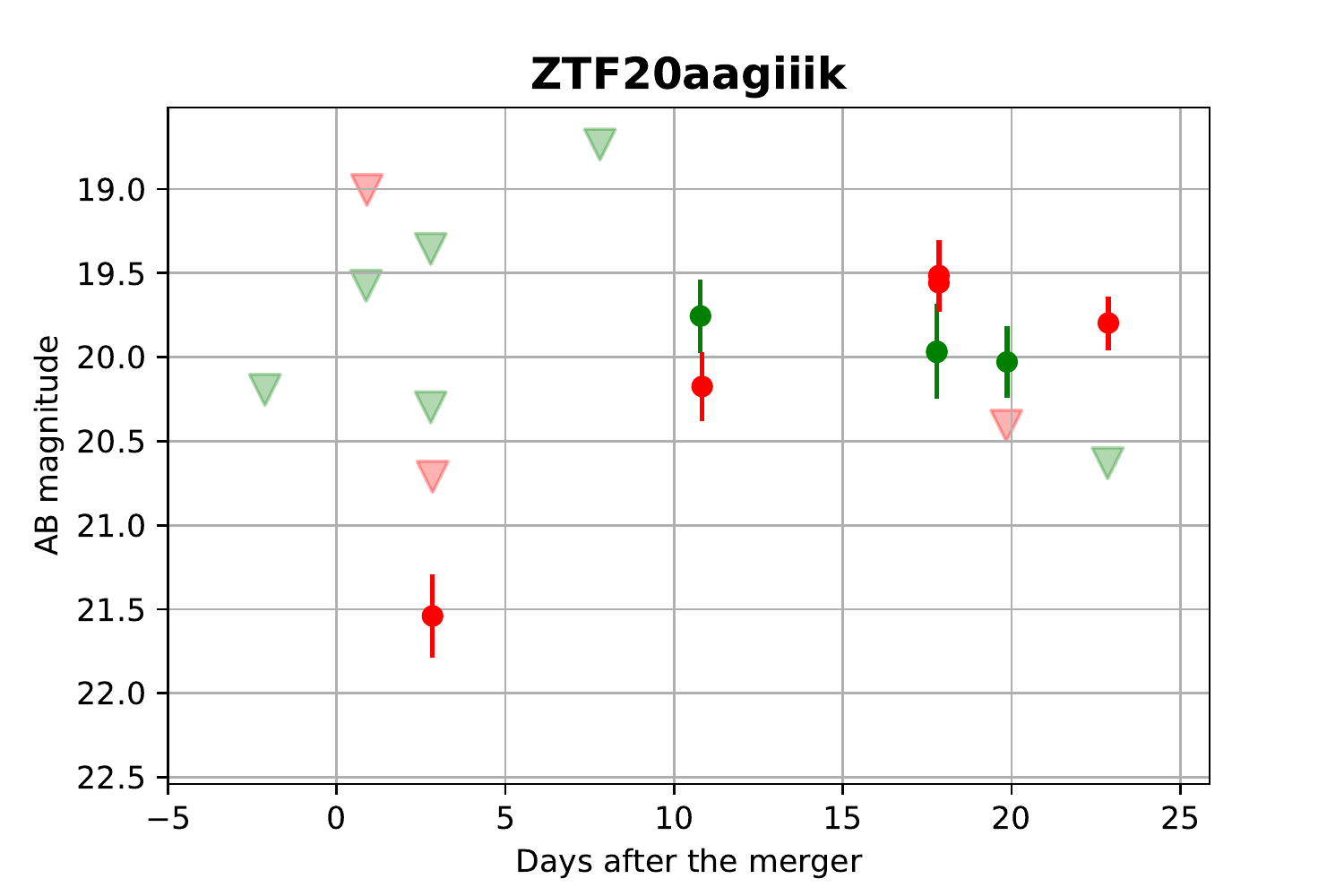}\includegraphics[width=2.25in]{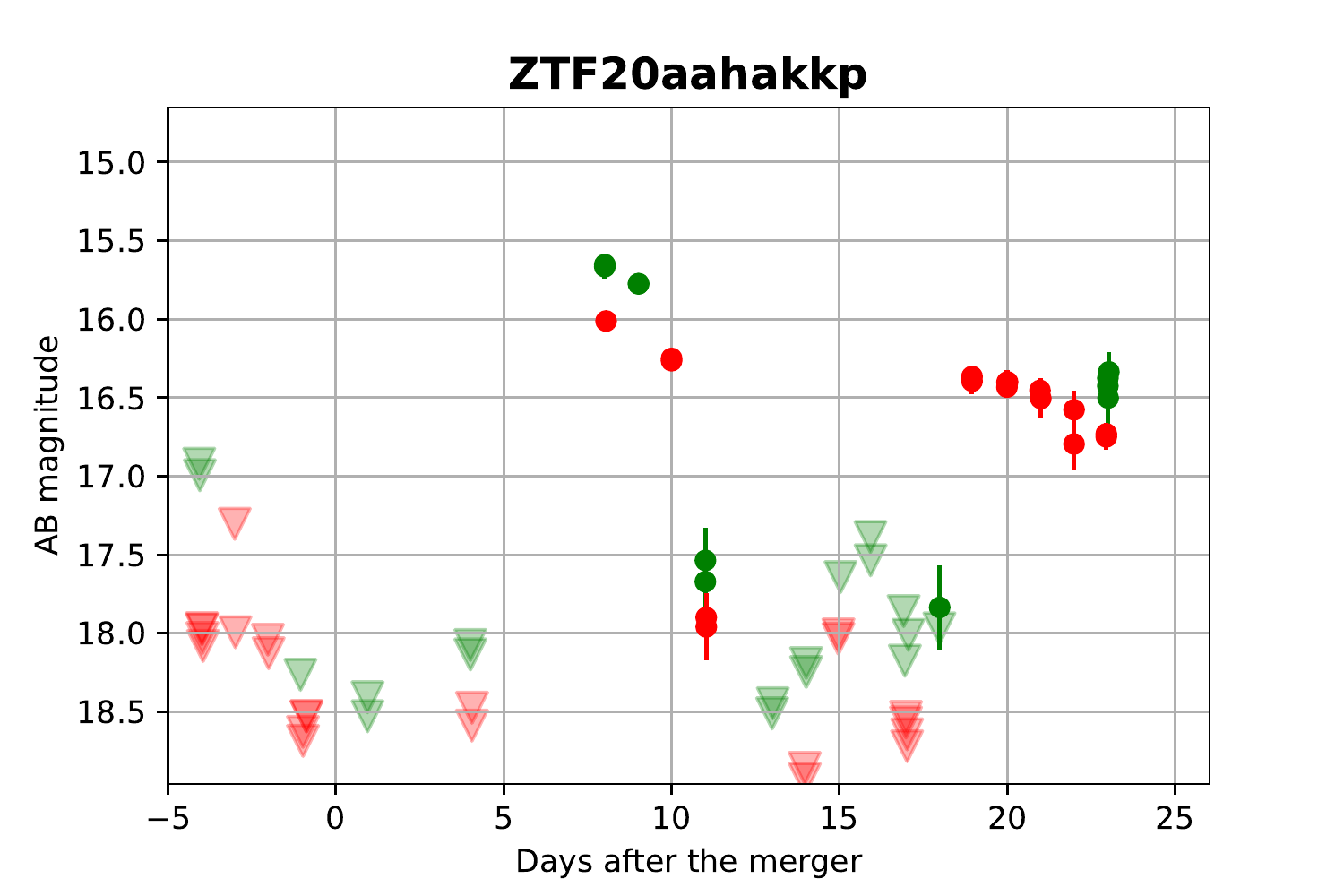}
    \includegraphics[width=2.25in]{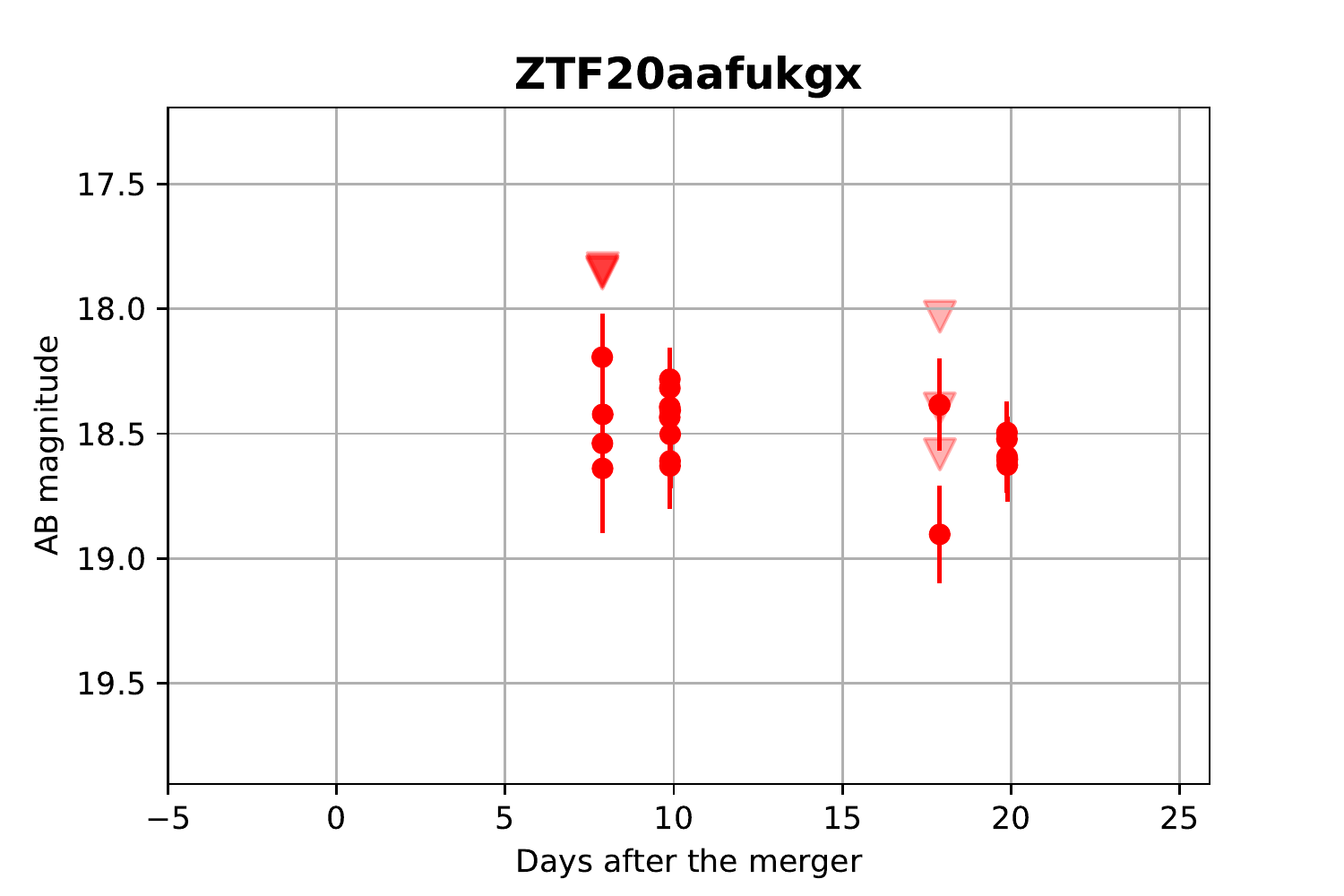}\includegraphics[width=2.25in]{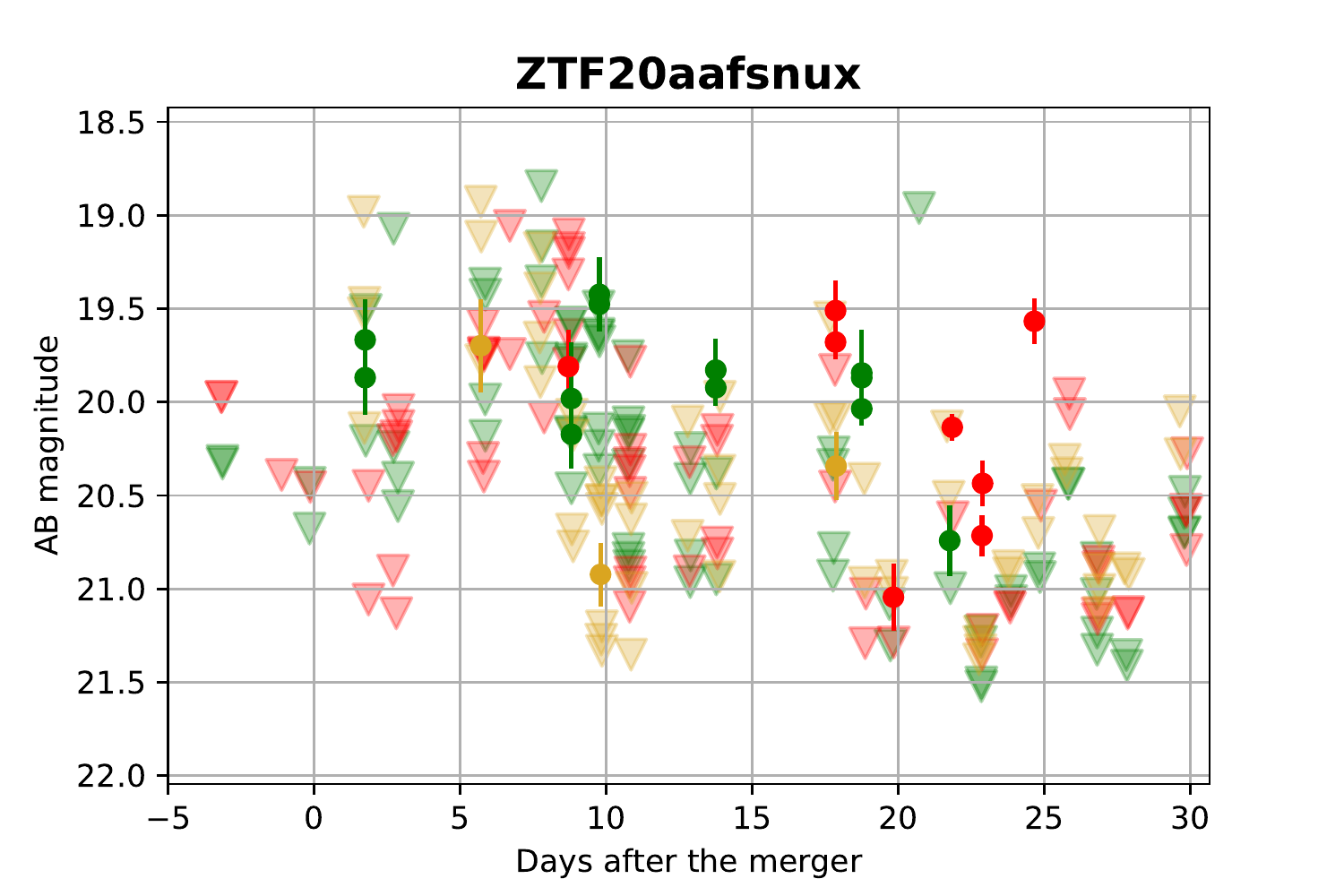}\includegraphics[width=2.25in]{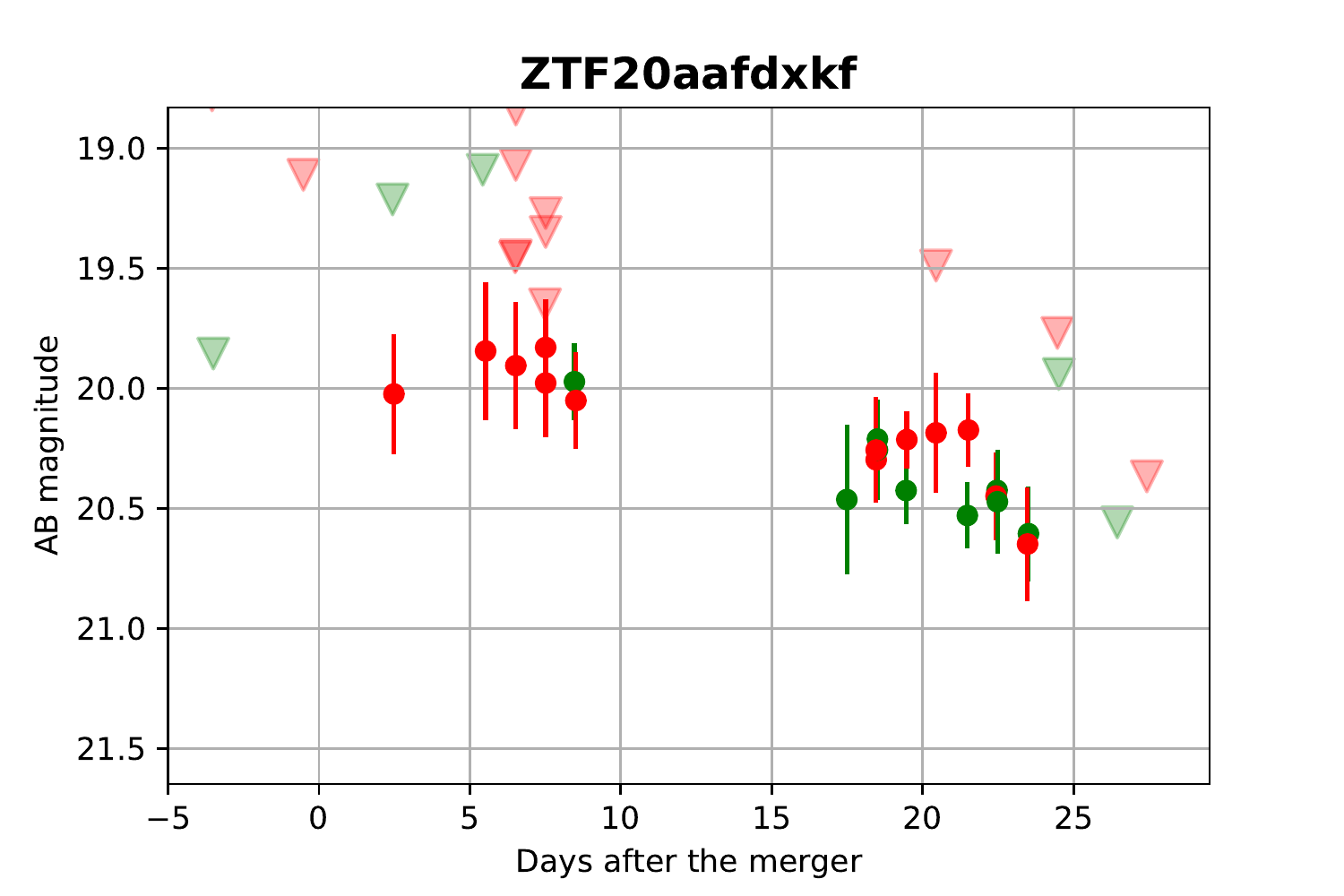}
    \includegraphics[width=2.25in]{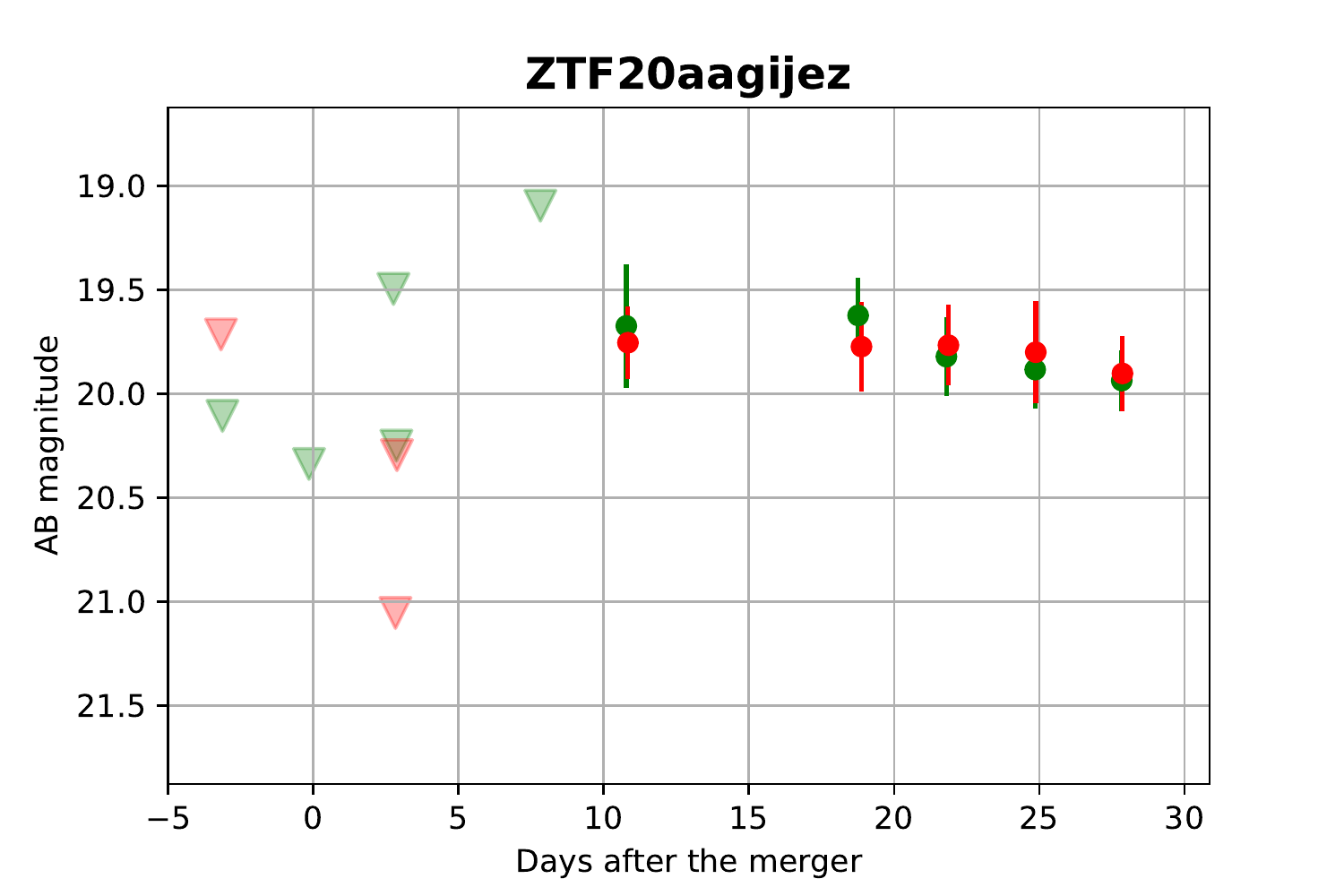}\includegraphics[width=2.25in]{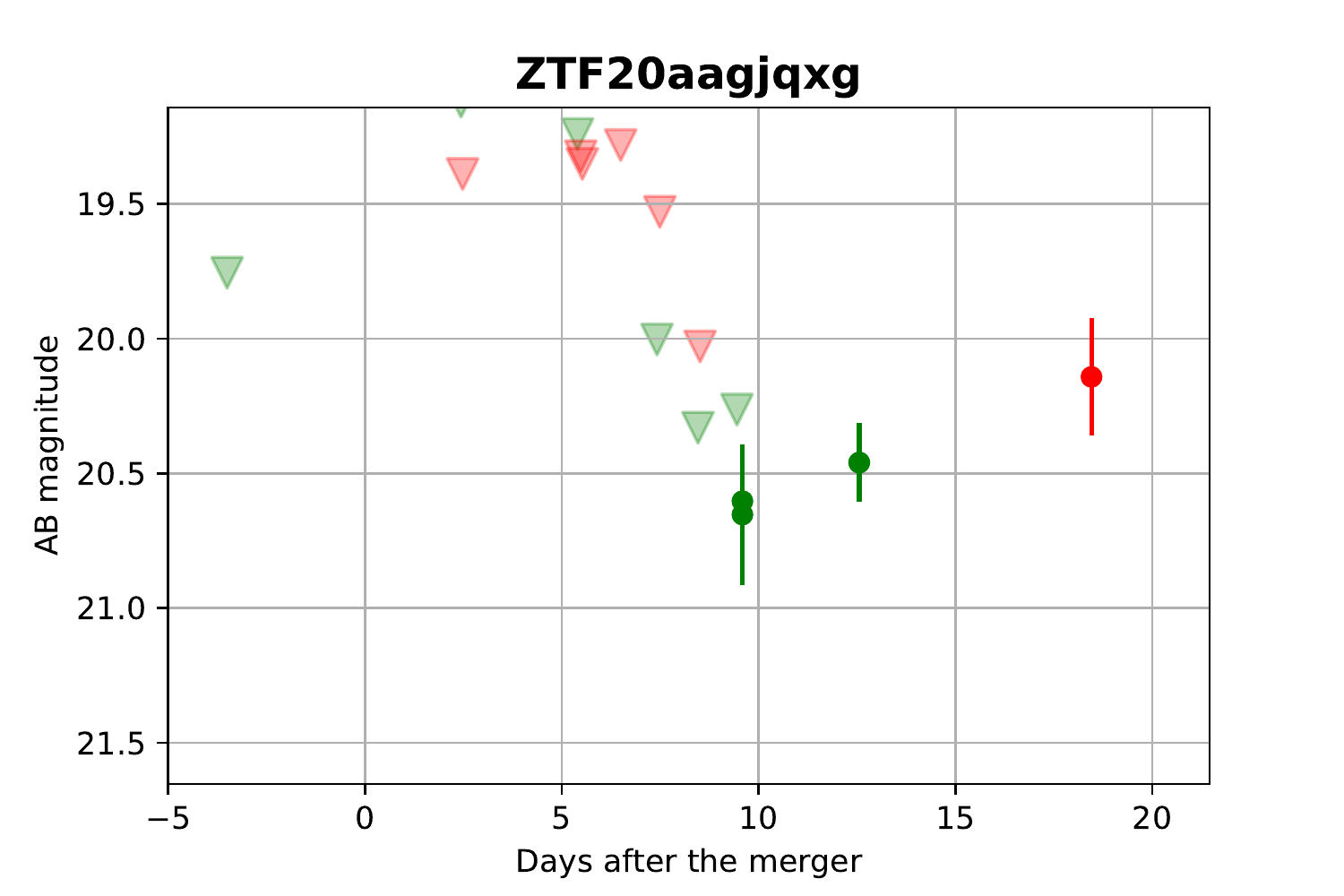}\includegraphics[width=2.25in]{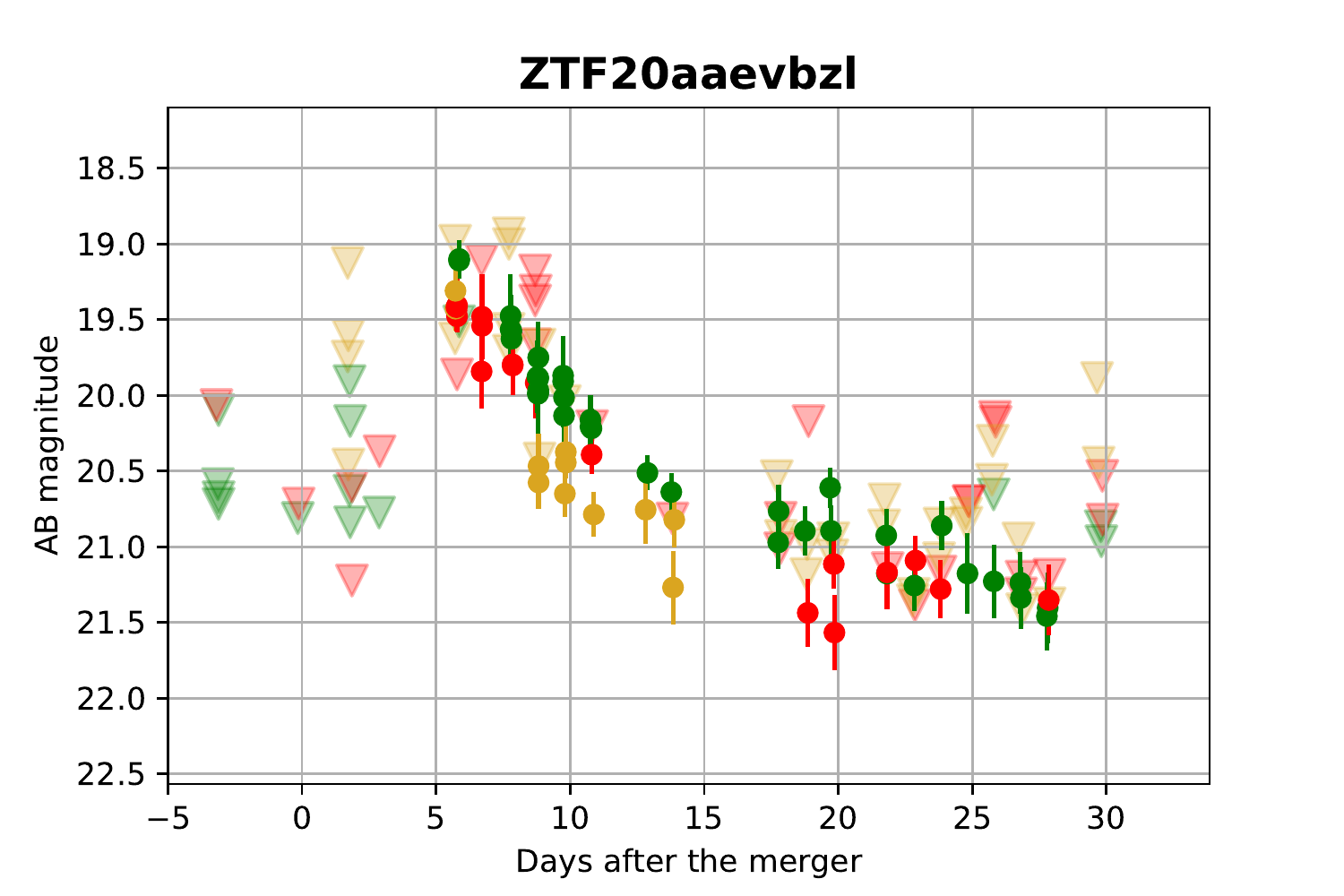}
    \includegraphics[width=2.25in]{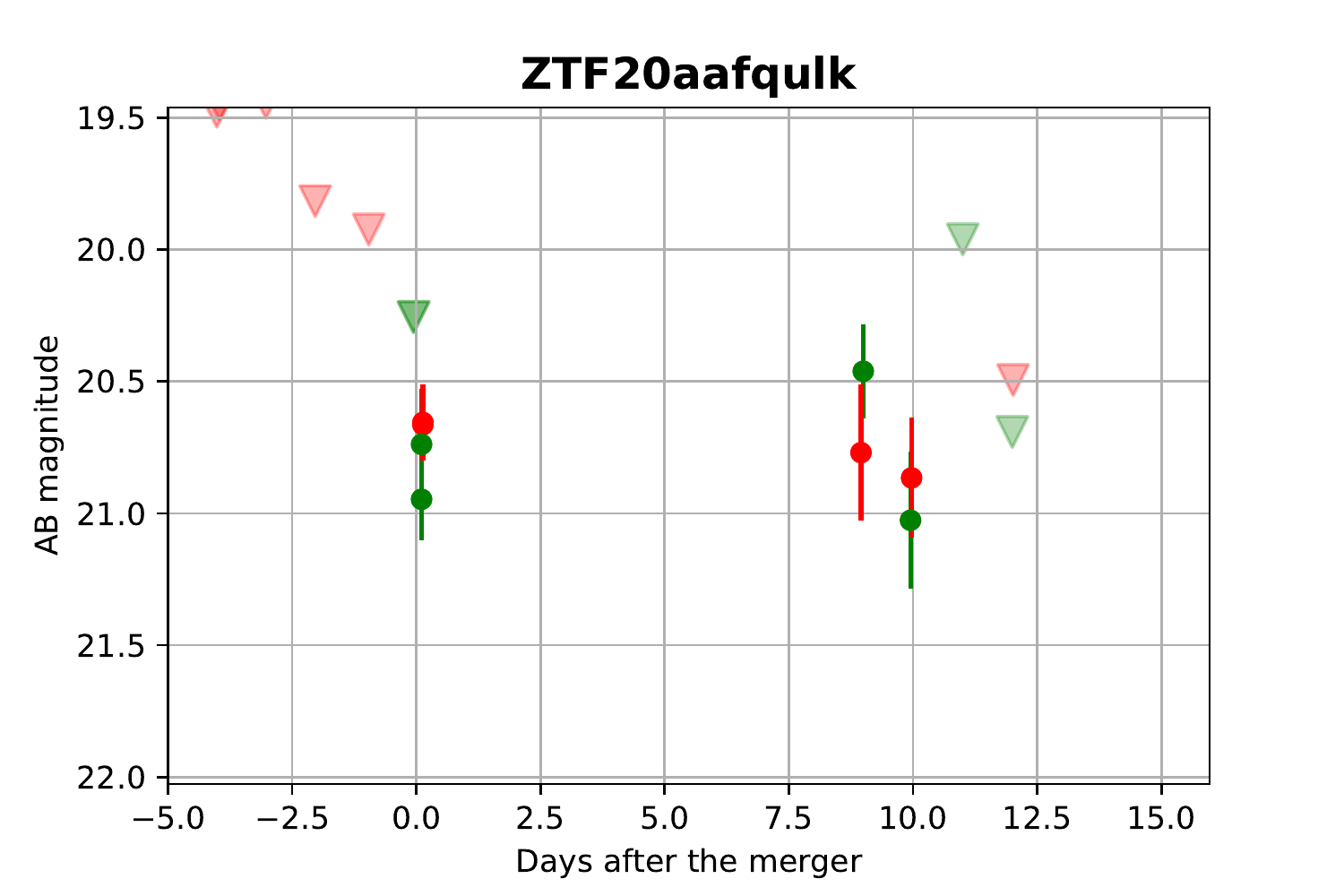}\includegraphics[width=2.25in]{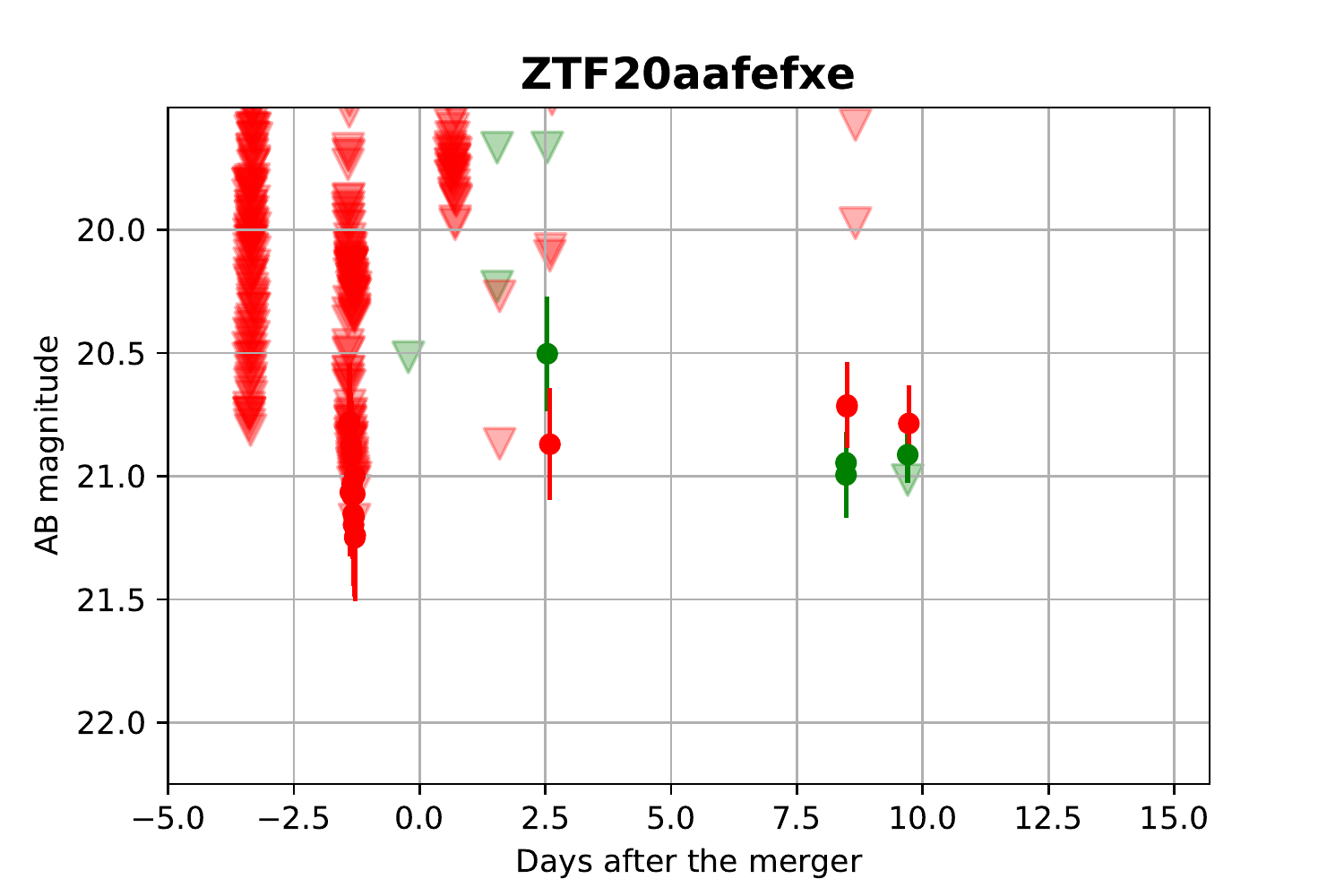}\includegraphics[width=2.25in]{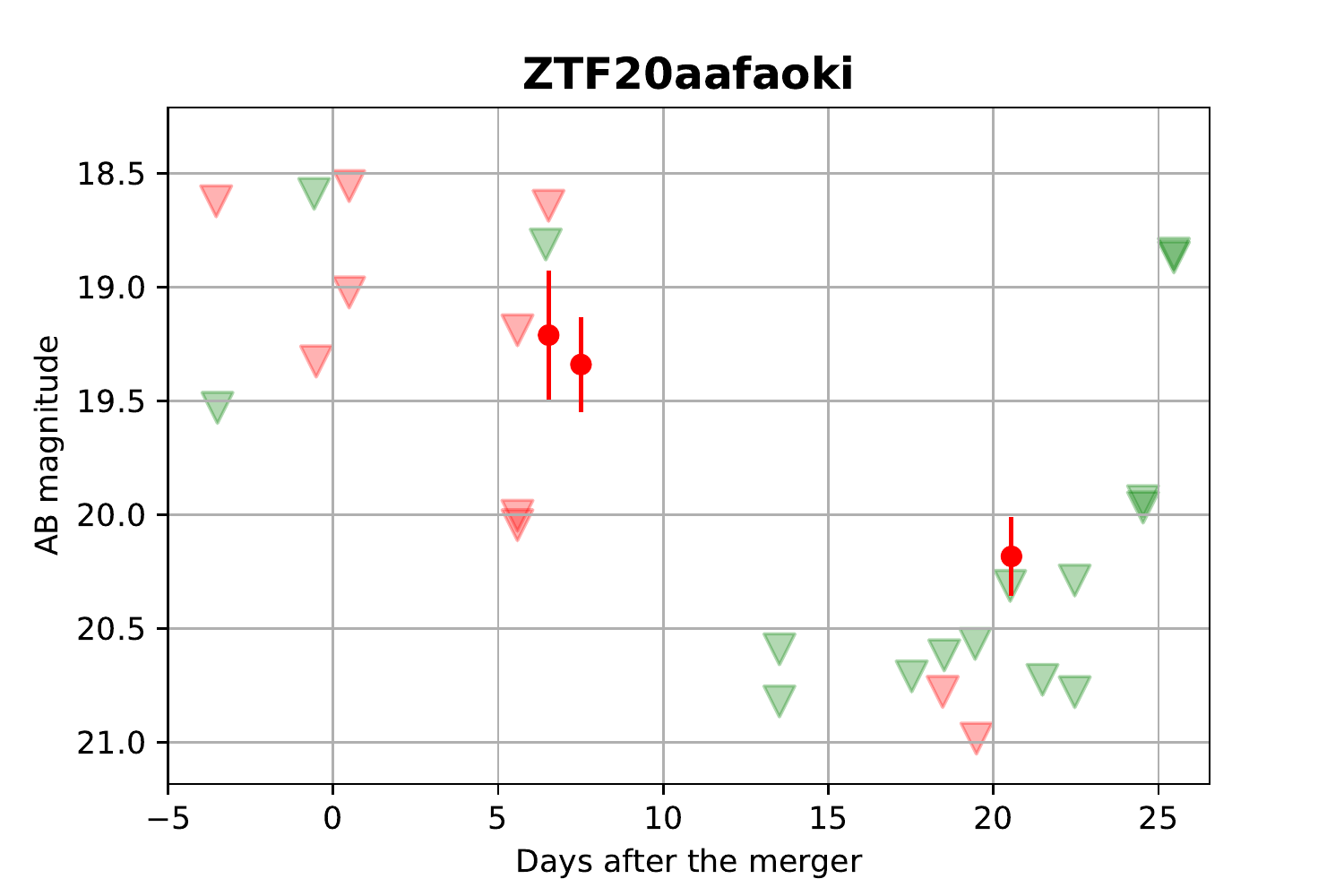}
    \caption{\textbf{Lightcurves for all objects ruled out photometrically.} In each panel, filled circles represent ZTF forced photometry and the photometry from the ZTF alert production pipeline, with error bars corresponding to 1-$\sigma$ uncertainties. Filled triangles display 5-$\sigma$ upper limits for non-detections. The $r$-, $g$-, and $i$-band data is presented in red, green and yellow respectively.}
    \label{fig:S200105ae_lcs}
\end{figure*}

All the transients are displayed in Supplementary Information Table~\ref{table:S200105ae_followup}; here we briefly describe each set, and show examples of light curves and cutouts for the most well-sampled, slowly photometrically evolving ones in \ref{fig:S200105ae_lcs}. For the candidates with spectroscopic redshifts, we compute their distance assuming Planck15 cosmological parameters and use them to estimate the source absolute magnitudes, which we include in the candidate descriptions. When vetting, we prioritized candidates whose distance fell within the 1$\sigma$ LIGO distance uncertainty for each event; however we did not reject any candidates on the basis of redshift.

The redshifts presented in this section come either from the spectra of the transient, z(s), or from the Photometric Redshifts for the Legacy Surveys (PRLS) catalog (Zhou et al. in prep.), which is based on Data Release 8 of DESI Legacy Imaging Surveys \citeNew{DeSh2019}, z(p). 

\subsubsection*{Spectroscopic Classification}

For this set of spectra, we quote the photometric phase at which the spectrum was taken when the photometry is well-sampled. In all other cases, we derive the spectroscopic phase of the transient using SNID \citeNew{BlTo2007} unless otherwise specified.  Most of the spectroscopic classifications were determined using SNID.

\textit{ZTF20aaertpj} - The first $r$- and $g$-band detections of this transient 3 days after the merger showed a red color $g-r$ = 0.4\,mag; it rapidly brightened 1\,mag to reach $g$ = 18.9 after 7 days. The Gran Telescopio Canarias (GTC) classified it as a Type Ib SN (z(s) =  0.026) on January 10th \citeNew{gcn26703} a few days before the ZTF lightcurve reached maximum light, implying an absolute magnitude of $-$15.9\,mag. This supernova is closer than the $-1\sigma$ LIGO distance.

\textit{ZTF20aaervoa} - This object was found 3 days after the merger at 20.74\,mag in $g$ band with a red color ($g-r=0.66$\,mag). This field was last observed 1.6 days before the merger. It showed a flat evolution over the first few days. Spectroscopic follow-up with GTC on January 10th classified it as a SN Type IIP  (z(s) = 0.046), $\sim$ 3\ days after maximum \citeNew{gcn26702} using SNID templates. This implied an absolute magnitude of $-$16.4\,mag in $r$ band. Its redshift is marginally consistent with the LIGO distance uncertainty, though it fell outside the 95\% confidence level of the LALInference skymap.

\textit{ZTF20aaervyn} - Its first detection was in the $g$ band ($g$ = 20.62\,mag), 3 days after the merger, which first showed a red color ($g-r=0.3$\,mag). This field was last visited 3 hours before the LVC alert. It was classified by GTC on Janunary 11th as a Type Ia SN, with z(s) = 0.1146 [ref.~\citeNew{gcn26702}], much farther than $+1\sigma$ LIGO distance. The spectroscopic phase corresponds to $\gtrsim$ 1 week before the lightcurve reached maximum light.

\textit{ZTF20aaerxsd} - Similarly, this region was visited 3 hours before the LVC alert and this candidate was first detected 3 days after the merger at $g$ = 20.27\,mag and showed a red color of $g-r$ = 0.37\,mag. The next couple of detections showed a quickly evolving transient, brightening $\sim 0.35$ mag/day. GTC spectroscopically classified it as a SN Type Ia (z(s) = 0.0533) on January 10th \citeNew{gcn26702}; concurrent photometry with ZTF indicates that the spectrum was taken $>12$ days before maximum.

\textit{ZTF20aaerqbx} - This transient was first detected in $g$-band at $g$ = 19.46\,mag 3 days after the merger. It faded 0.5\,mag over the first 8 days and was classified by GTC on January 11th as a Type IIP SN (z(s) = 0.098) at 5 days before maximum, using SNID \citeNew{gcn26703}. Its redshift places it outside of the LIGO volume.

\textit{ZTF20aafanxk} - This candidate was detected at $r$ = 18.52\,mag, 6 days after the merger with galactic latitude $<15^\circ$ and offset by 7\arcsec\,from a possible host \citeNew{gcn26810}; it faded 0.3\,mag in the $r$-band the first 10 days and a spectrum taken with the P60 SEDM spectrograph revealed its classification to be a SN Ia at z(s) = 0.103, too far to be consistent with the LIGO distance.

\textit{ZTF20aafujqk} - Offset by 2.26\arcsec\,from the center of a large spiral galaxy host \citeNew{gcn26810}, ZTF20aafujqk was detected in $r$-band during serendipitous observations 10 days after the merger, and later followed up with SEDM photometry in $g$- and $i$- bands, which showed a steadily declining lightcurve.  SEDM spectroscopy showed that it was also a SN Ia at z(s) = 0.06, consistent with LIGO distance uncertainties.

\textit{ZTF20aaevbzl} - This region was last observed 3 hours before the LVC alert. ZTF20aaevbzl was detected six days after the merger \citeNew{gcn26810}, this candidate was selected for its atypical rapid decline in its lightcurve in $r$- and $g$-bands. This hostless transient faded 1.1\,mag in 5 days in the $g$-band. We obtained a spectrum of ZTF20aaevbzl with P200+DBSP, whose H$\alpha$ feature at z(s) = 0 amidst a blue, mostly featureless spectrum indicates that it is a galactic cataclysmic variable (See Figure \ref{fig:spectra}). Further follow-up with SEDM and LCO showed that the transient was consistently fading at 0.18 magnitudes per day in the g- band. 

\subsubsection*{(Slow) Photometric Evolution}

\begin{figure}[t]
 \includegraphics[width=\columnwidth]{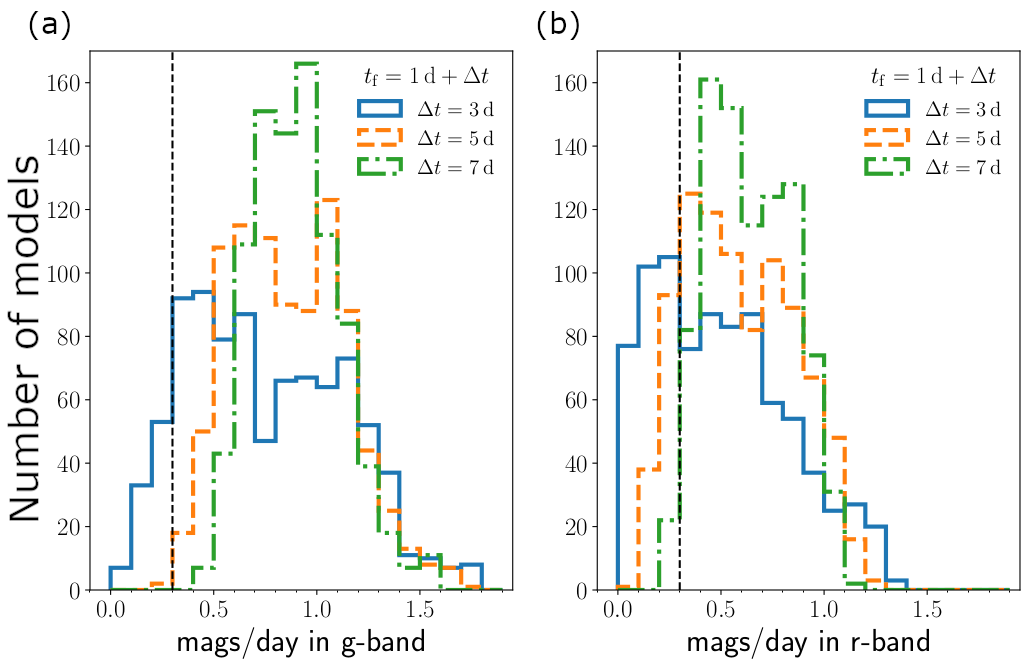}
  \caption{\textbf{Plot of the decay rate (mag/day) in $g$-band (a) and $r$-band (b) for all the ejecta masses and viewing angles of the modeled grid presented in Section~\ref{sec:kilonovae}.} Blue histograms are for time windows from 1 to 4 days after merger ($\Delta t=3$\,days), orange from 1 to 6 days ($\Delta t=5$\,days), green from 1 to 8 days ($\Delta t=7$\,days). In general, $96\,\%$ of models show faster decay than 0.3\,mags/day (dashed vertical line) in $g$-band, while $82\,\%$ of models show faster decay than 0.3\,mags/day in $r$-band. The more slowly fading models are the higher mass ones. Particularly, our threshold was chosen based on the 7 days baseline, as all the candidates meet that requirement.
  }
 \label{fig:evolution}
\end{figure} 

As mentioned above, we deem candidates to be slowly evolving by checking whether their rise or decay rate is faster than our photometric cut of $<|0.3|$ mag/day. We justify this cut based on \ref{fig:evolution}, a histogram of the evolution rates of KNe from NSBH mergers, which shows that over a baseline of $\gtrsim$1 week, which is the case for our candidates, nearly all KN model lightcurves evolve faster than this cut in both $g$- and $r$-bands. The decline rate is determined using the photometric band with the longest available baseline. It is calculated by getting the ratio between the $\Delta m$ and the length of that baseline ($\Delta t$), from the candidate's peak to its last detection.  This cut does exclude from our analysis a small part of the physically acceptable parameter space of NSBH binaries (see \ref{fig:pelim}), though it significantly reduced the number of false-positive transients. It should thus be seen as a trade-off between parameter space coverage and the cost of EM follow-up that result in a small and known bias in our search.

\textit{ZTF20aafduvt} - The field where this transient lies was observed 12 hours before the LVC alert, and it was detected six days after the merger in $r$- and $g$- bands \citeNew{gcn26810}, offset from a possible host at z(p) = $0.21 \pm 0.02$ by 51kpc, this candidate faded 0.1\,mag in the $g$-band during the first 9 days after the discovery. The photometric redshift places this transient at an absolute magnitude of $M$ = -21\,mag.

\textit{ZTF20aaflndh} - With its last non-detection 12 hours before the GW alert, ZTF20aaflndh was first detected 10 days after the merger. This source is located 0.8\arcsec\,from the center of an apparently small galaxy \citeNew{gcn26810} and evolved photometrically 
to resemble a Type Ia SN light curve; it faded in the $r$-band by 0.17\,mag in 17 days. Furthermore, the photo-z of the host galaxy is z(p) = 0.091 $\pm$ 0.023 which puts the transient at an absolute magnitude of $M$ = -19.06\,mag, consistent with a Type Ia SN.

\textit{ZTF20aaexpwt} - This candidate was first detected one week post-merger, and was one of several hostless candidates identified in a low galactic latitude (b$_{\rm gal} < 15^{\circ}$) field \citeNew{gcn26810}. The last non-detection was 5 hours before the LVC alert. Its evolution over the next seven days was 0.12\,mag/day in the $r$-band, marked by a declining lightcurve.

\textit{ZTF20aafukgx} - Offset from a potential bright host by 3.85\arcsec, at low galactic latitude \citeNew{gcn26810}, this candidate was detected at $r$ = 18.4 ten days after the merger but remained flat within error-bars over the next ten days of observations. 

\textit{ZTF20aagijez} - First detected 11 days post-merger, this candidate, offset 3.15\arcsec\,from the nucleus of a star-forming galaxy at z(s) = 0.061 [ref.~\citeNew{gcn26810}], exhibited a flat lightcurve for more than 10 days and it was still detectable after 40 days; it photometrically resembles a SN light curve. The spectroscopic host redshift implies an absolute magnitude of $M$ = -17.6\,mag. The last visit to the field where this transient lies was 3.6 hours before the GW alert. 

\textit{ZTF20aagiiik} - This field was last visited 2 days before the LVC alert. We identified ZTF20aagiiik as a candidate of interest due to its rapid rise in $r$-band after being detected 11 days after the merger; it is offset by 5.79\arcsec\,from a potential spiral galaxy host \citeNew{gcn26810}. However, it only faded 0.4\,mag in 12 days. Additionally, at the redshift of the potential host galaxy (z(s) = 0.13, separated by 5.25\arcsec) the absolute magnitude ($M$ = -19.24\,mag) is consistent with a Type Ia SN. 

\textit{ZTF20aafdxkf} - Detected just three days after the merger, this hostless candidate exhibited a rise in $r$-band over the first three days \citeNew{gcn26810}, but its declining $g$-band photometry showed it to be too slow to be a KN. It only faded 0.5\,mag in the $g$-band during the first 14 days. The last non-detection was 12 hours before the LVC alert.

\textit{ZTF20aagiipi} - Offset by 27 kpc from a potential faint host at z(p) = $0.388 \pm 0.016$, this candidate seemed to be rising when it was detected in the first 11 days after merger. Supplemented with SEDM photometry, its lightcurve closely resembles that of a typical Type Ia supernova, which at the redshift of the host would peak at $M$ = -21.6\,mag. This field was last observed 3.6 hrs before the LVC alert. 

\textit{ZTF20aafsnux} - A hostless candidate, ZTF20aafsnux appeared to be declining gradually based on its first two $g$-band detections two and nine days after the merger. Close monitoring revealed that the source was fluctuating between $g\sim$ 19.0--20.0\,mag over a period of 17 days. This region was last visited 3 hours before the GW alert. 

\textit{ZTF20aaertil} - This candidate was first detected three days after the merger; it was located 0.2\arcsec from the nucleus of a faint galaxy host and appeared to be rising in $g$-band \citeNew{gcn26810}. Our spectrum of the host galaxy with DBSP on Jan 18th demonstrated that the galaxy, at z(s) = 0.093, was outside the one-sigma distance uncertainty for S200105ae; furthermore, in 40 days, it faded only 0.5\,mag in the $r$-band. The absolute magnitude at this host redshift is $M$ = -18.5\,mag. We show the lightcurve and $r$-band cutouts for this transient in \ref{fig:lc_transients}. The last non-detection in this field was 3 hours before the LVC alert.

\textit{ZTF20aafksha} -  This last non-detection for this transient was 1.2 days before the GW alert. We discovered this candidate nine days after the merger, offset by 7.92\arcsec\,from a possible spiral galaxy host at z(s) = 0.167 at $g$ = 20.06\,mag [ref.~\citeNew{gcn26810}], corresponding to an absolute magnitude of about $-$19.6\,mag. The steadily declining lightcurve post-peak in both $g$-band and $r$-band, 0.7\,mag in $g$-band during the first 19 days, and the bright absolute magnitude, suggests that the candidate is a SN Ia. We display this candidate in \ref{fig:lc_transients}.

\textit{ZTF20aagjemb} - First detected 3 days after merger, this nuclear candidate rose by one magnitude over the course of 5 days in $g$-band \citeNew{gcn26810}. After tracking its evolution over 20 days time, the lightcurve seems to exhibits a SN-like rise and decline. It presents a slowly-evolving lightcurve, only fading 0.1\,mag in the $r$-band during the twenty days. This candidate is also displayed in \ref{fig:lc_transients}. The transient is located in a host with a z(p) = 0.21 $\pm$ 0.06, separated by 6 kpc, implying an absolute magnitude $M$ = -19.24\,mag. The last non-detection in this region was 3 hours before the LVC alert.

\textit{ZTF20aafefxe} - This candidate's two detections in $r$-band suggest fading behaviour, but subsequently the source has not been detected by the nominal survey observations \citeNew{gcn26810}. The last non-detection in this region was 5 hours before the LVC alert. The first detection was 9 days after the merger, and there may be a faint host separated by 41 kpc from the transient with z(p) = 0.09 $\pm$ 0.05, indicating a luminosity of $M$ = $-$17.2\,mag. Forced photometry revealed that it had only evolved 0.16 mags in 11 days in the $g$-band, placing it clearly into the category of slow evolvers. 

\textit{ZTF20aafaoki} - The last non-detection in this region was 12 hours before the LVC alert. This candidate had two $r$-band detections at 19.2\,mag, but had faded below 21.4\,mag just 5 days later \citeNew{gcn26810}. Our images taken with KPED do not show any transient or background source up to $g>$ 19.55\,mag 6 days after the discovery. Similarly, our LCO follow-up observations showed that 8 days after the discovery, the transient is not detected and there is no visible source at the corresponding coordinate up to $g >$ 20.25\,mag and $r >$ 21.6\,mag. Our last LCO observations, obtained 72 days after the discovery, show no transient up to $g>22.10$\,mag. However, after running forced photometry at the transient position, we find a detection 14 days after the initial discovery at $r$ = 21.2\,mag, implying re-brightening of the transient after the non-detection upper limits, or very slow evolution.


\subsubsection*{Stellar}

\textit{ZTF20aafexle} - This particular region was observed serendipitously 1 hour before the LVC alert. After its initial detection 8 days after the merger, it brightened by nearly one magnitude over four days but returned to its original brightness after 5 days \citeNew{gcn26810}. We posit that it may be stellar due to the PS1 detections at the source position. Additionally, its evolution over the first 10 days after the discovery is only 0.3\,mag in the $r$-band. 


\subsubsection*{Slow-moving asteroids}

\begin{figure}[!htb]
  \begin{minipage}[b]{\linewidth}
    \centering
    \includegraphics[width=\linewidth]{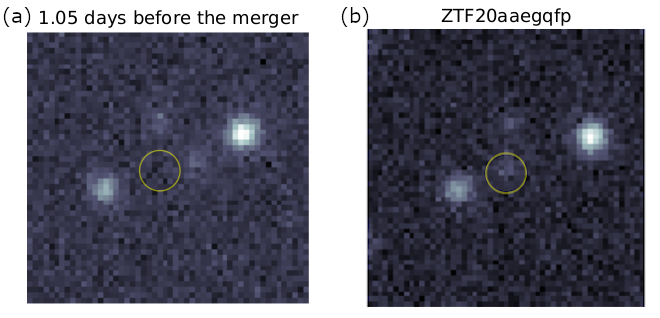}

  \end{minipage}
  
\caption{\textbf{ZTF $r$-band cutouts of the slow moving asteroid ZTF20aaegqfp.} The yellow  circles show the position of the ZTF candidate in both cutouts. Panel (a) shows a cutout of the region one day before the trigger. There, it is possible to see a source to the right of ZTF20aaegqfp position, marked with a yellow circle. This source was located at 7.3\,\arcsec\, from our candidate. Panel (b) shows the discovery image of our candidate ZTF20aaegqfp, which is located within the circle. The cutouts are 0.7 sq. arcmin and north and east are up and to the left respectively.  }\label{fig:ast_ZTF20aaegqfp}
\end{figure}

\textit{ZTF20aaegqfp} - We detected this hostless candidate a day after the merger in $r$ band. The last non-detection of this transient was 5 hours before the GW alert. Our pipelines identified it as a fast-evolving transient due to its rise by more than 0.5\,mag over the course of the night; subsequently, it was not detected in any our serendipitous observations. We find non-physical upper limits interspersed with detections, suggesting that the photometry for this transient may not be reliable. Using the Kowalski infrastructure, we queried for alerts in the vicinity of the transient (around 25\arcsec) and found 13 alerts, the oldest of which was $\sim 4$ days before the trigger, which showed a moving object across the field alerts (see \ref{fig:ast_ZTF20aaegqfp}).

\subsection{S200115j candidates}\label{section:15jcandidates}

In this subsection, we provide brief descriptions of candidates identified within the skymap of S200115j. Most of our candidates were identified during the serendipitous coverage of the map. Some of our transients were discovered within ZTF Uniform Depth Survey (ZUDS; Goldstein et al., in prep) a dedicated survey for catching high-redshift SNe by acquiring and stacking images to achieve greater depth compared to the nominal survey. Intrinsically faint transients ($m_\mathrm{AB} \sim -16$\,mag) discovered in these fields are more likely to be at redshifts consistent with the distance of this event ($340 \pm 79$\, Mpc).

The relevant candidates circulated by the GROWTH collaboration \citeNew{gcn26767} were found on the first night of observations. Weather issues affected systematic follow-up in the following days; nevertheless, a later deeper search led to more candidates found to be temporally and spatially consistent, which we report here.
Additionally, candidates from Ref.~\citeNew{gcn26798} were cross-matched with the ZTF database in order to temporally constrain the transients. Only S200115j$\textunderscore$X136 \citeNew{gcn26798} had an optical counterpart we could identify, ZTF20aafapey, with a flaring AGN \citeNew{gcn26863}.

Every candidate that was found in the region of interest is listed in Supplementary Information Table~\ref{table:S200115j_followup}.

\subsubsection*{Spectroscopic Classification}

\textit{ZTF20aafqpum} - This transient is located at the edge of a host galaxy at photz $=0.12 \pm 0.03$ [ref.~\citeNew{gcn26767}].  The region was last observed 1 hour before the LVC trigger and the transient.
Follow-up with the Liverpool telescope in $r$- and $i$-bands showed this candidate to be red, with $g-r\sim0.5$\,mag. This transient was then spectroscopically classified by ePESSTO+ as a SN Ia 91-bg, at z(s) = 0.09 [ref.~\citeNew{ScIr2020}], placing it at an absolute magnitude of $M$ = $-$17.3\,mag.

\subsubsection*{(Slow) Photometric Evolution}

\begin{figure*}[!htb]
  \begin{minipage}[b]{0.48\linewidth}
    \centering
    \includegraphics[width=\linewidth]{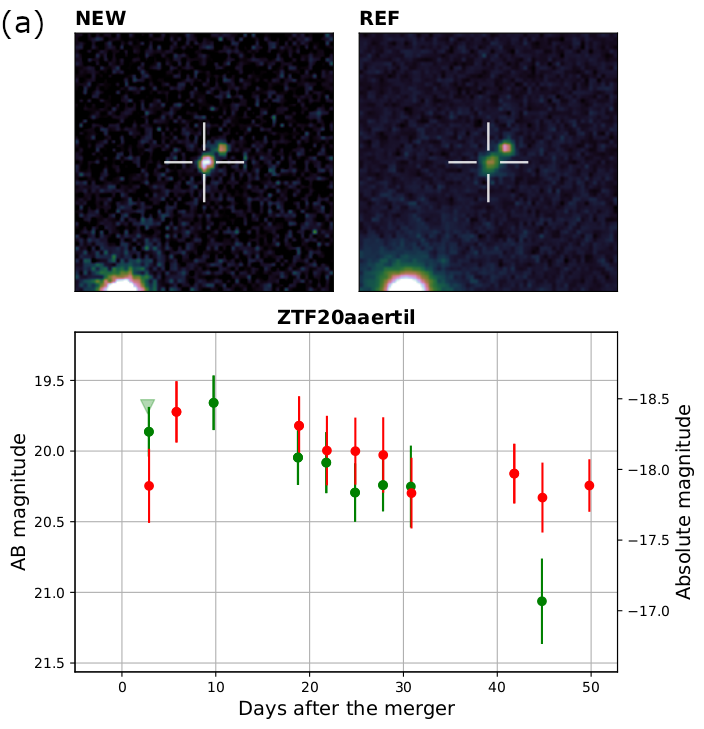}

  \end{minipage}
  \begin{minipage}[b]{0.48\linewidth}
    \centering
    \includegraphics[width=\linewidth]{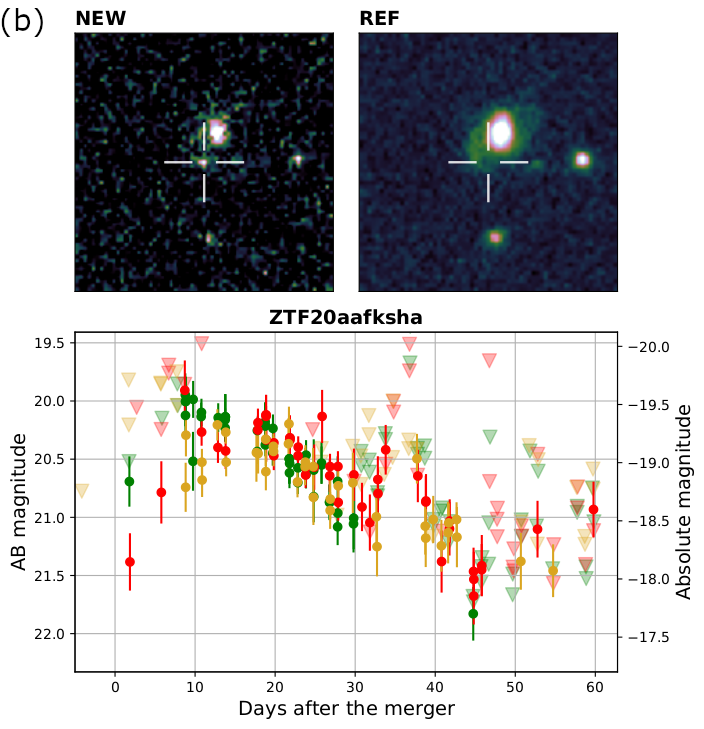}
  \end{minipage} 
  \begin{minipage}[b]{0.48\linewidth}
    \centering
    \includegraphics[width=\linewidth]{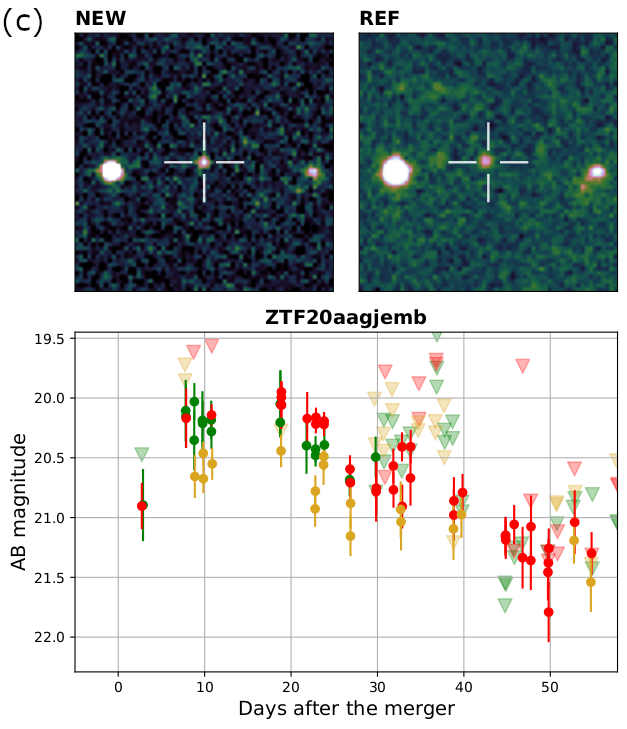}
  \end{minipage}
  \begin{minipage}[b]{0.48\linewidth}
    \centering
    \includegraphics[width=\linewidth]{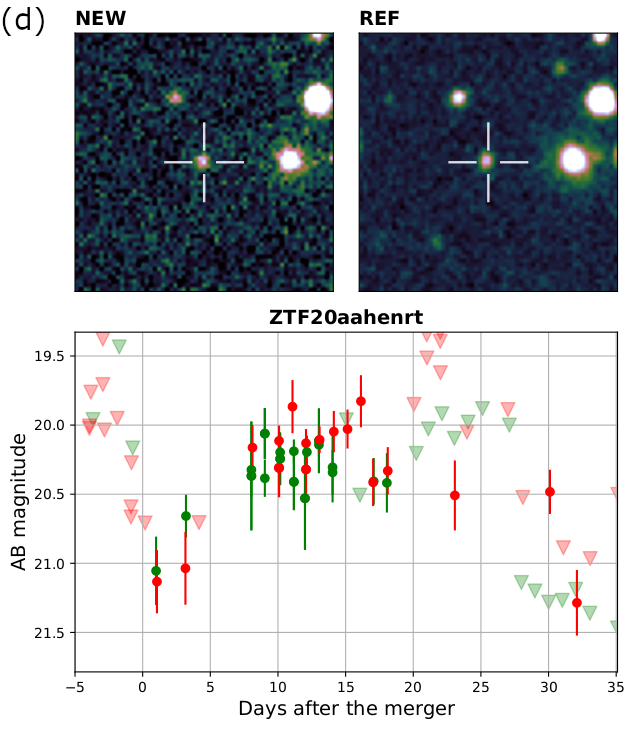}
  \end{minipage} 
\caption{\textbf{Lightcurves and $r$-band cutouts for a subset of the most well-sampled lightcurves for ZTF candidates that were ruled out photometrically.}  Colors were used to represent the different bands: green, red and yellow for $g$-, $r$- and $i$- bands. The triangles in the lightcurve represent upper limits and filled circles are the detected magnitudes of the object. On each panel, the left cutout is the ZTF discovery image and the right one is the corresponding ZTF reference image. The transient is marked with a cross and the size of the cutouts is 0.7 sq. arcmin with north being up and east to the left. The candidates highlighted here are as follows: (a) ZTF20aaertil, (b) ZTF20aafksha, (c) ZTF20aagjemb, and (d) ZTF20aahenrt.}
\label{fig:lc_transients}
\end{figure*}

\textit{ZTF20aahenrt} - This candidate, detected during our serendipitous search 3 days after the merger, is separated from a galaxy host by 8.8 kpc at z(p) = 0.16 $\pm$ 0.04, giving it an absolute magnitude of $M$ = $-$15.6\,mag. We monitored the transient after its initial rise in $g$-band, but over 12 days the candidate lightcurve exhibits very flat evolution, rising by 0.14\,mag in 7 days. We highlight it in \ref{fig:lc_transients} as an example of a very slowly evolving transient identified in our searches. This field was serendipitously observed 30 min before the LVC alert.

\textit{ZTF20aagjqxg} - We selected this hostless candidate during our scanning due to its faint $g$-band detection at $g$ = 20.65\,mag and subsequent rise three days after the initial detection two hours after the merger; its detection 11 days later in the $r$-band suggests that it was rising or reddening at a rate of $<$ 0.1\,mag/day. This field was last observed 3.5 days before the LVC alert.

\textit{ZTF20aahakkp} - This hostless transient was first detected eight days after the merger in $g$ = 15.67\,mag and $r$ = 16.01\,mag. The last non- detection of this transient was 20 hours before the issue of the LVC alert. While the transient seems to be rapidly fading over the course of a day from $r$ = 16.26\,mag to $r$ = 17.9\,mag, this detection is likely affected by poor weather and bad seeing on that day (seeing 4\arcsec). 20 days later, the lightcurve is near the original detection magnitude, and exhibits a slow fade since then.


\textit{ZTF20aafqulk} - This region was last observed 1 hour before the issue of the GW alert. This source was detected 2.5 hours after the merger in $g$-band and 43 minutes later in $r$-band, with a blue color ($g$-$r$ = 0.2\,mag).
The candidate is offset by 0.3\arcsec\,from a potential host galaxy at a photometric redshift of z(p) = 0.27 $\pm$ 0.04 [ref.~\citeNew{gcn26767}]. Our P60+SEDM spectrum does not offer a clear classification, but we detect a source in our LCO images 5 days after its discovery with $r = 20.16 \pm 0.1$\,mag. When running forced photometry, we find a detection in the $r$-band 89 days before the trigger, definitively ruling out its association with the GW event. Furthermore, the lightcurve appears nearly flat in the $r$-band over the course of 10 days. 


\subsubsection*{Slow-moving asteroids}

Solar System asteroids located in the proximity of the stationary points located at $\sim$ 60$^\circ$ from opposition and low ecliptic latitude \citeNew{Green1985} have slow, $\lesssim$ 1\arcsec/h sky motions \citeNew{Jedicke2016}. 

\textit{ZTF20aafqvyc} - This was first detected as a hostless candidate 2.5 hours after the merger in $g$-band, followed by a detection in $r$-band just 49 minutes later \citeNew{gcn26767}. Due to the transient being faint at $g$ = 20.39\,mag, with a $g-r$ color of 0.34\,mag, we pursued follow-up with P200+WIRC on 2020-01-18 with NIR non-detections down to $J>$ 21.5\,mag and $K_{s}>$ 20.9\,mag [ref.~\citeNew{gcn26814}] and LCO on 2020-01-19 with optical non-detections down to $g>$ 22.6\,mag, $r>$ 21.8\,mag and $i>$ 20.9\,mag [ref.~\citeNew{gcn26817}]. Follow-up reported with AZT-33IK telescope of Sayan observatory (Mondy) revealed non-detections just 13 hours and one day after the merger, down to upper limits of 21.6\,mag and 22.1\,mag in the $r$-band, suggesting that the source could be fast-fading, if astrophysical \citeNew{gcn26819}. Finally, we conducted follow-up with Gemini GMOS-N, detecting no source down to an upper limit of $g>$ 24.5\,mag [ref.~\citeNew{gcn26822}]. Based on the puzzling non-detections, we investigated the possibilities that it could be an artifact or that it was a moving object. Close inspection of the images taken with the Liverpool Telescope, 12.9 hours after the merger in $g$- and $r$-bands clearly demonstrated that the object had shifted position in the image with a slow angular rate of motion consistent with being an asteroid with an opposition-centric location of $\pm$60$^\circ$ near the evening sky stationary point.

\section{Ejecta mass and binary parameter constraints -- Implications and caveats}

To further illustrate what we could learn from sufficiently deep observations, we consider potential constraints on the parameters of the NSBH binary powering S200105ae. We assume that the source was located at $283\,{\rm Mpc}$, and seen face-on. For the deepest fields reported here, we have seen that this implies $M_{\rm ej,dyn}\lesssim 0.02\,M_\odot$ and $M_{\rm ej,pm}\lesssim 0.04\,M_\odot$. Using semi-analytical formulae calibrated to the results of numerical simulations, we can estimate $M_{\rm ej,dyn}$ and $M_{\rm ej,pm}$ as functions of the mass ratio of the binary ($Q=M_{\rm BH}/M_{\rm NS}$), the component of the dimensionless black hole spin aligned with the orbital angular momentum ($\chi$), and the neutron star compactness ($C_{\rm NS}=\frac{GM_{\rm NS}}{R_{\rm NS}c^2}$) (see also~Refs.~\citeNew{CoDi2018,CoDi2018b,CoDi2019b,Andreoni2020,DiCo2020,CoDi2020}). We compute $M_{\rm ej,pm}$ using Ref.~\citeNew{Foucart:2018rjc}, and $M_{\rm rem}$ using Ref.~\citeNew{Kruger:2020gig}, which are based on, respectively, the work of Ref.~\citeNew{Foucart2012} and Ref.~\citeNew{KaKy2016}. As Ref.~\citeNew{Foucart:2018rjc} only predicts the total mass remaining outside of the BH after merger, $M_{\rm rem}$, we estimate $M_{\rm ej,pm}=f_{\rm rem}(M_{\rm rem}-M_{\rm ej,dyn})$,with $f_{\rm rem}\sim 0.15-0.5$ the fraction of the remnant accretion disk that is ejected in the form of disk winds~\citeNew{Christie:2019lim}. The results are shown in \ref{fig:pe}, expressed as the maximum BH spin compatible with the assumed mass constraints. We show results for $f_{\rm rem}=0.15$ and $f_{\rm rem}=0.5$, to illustrate the dependence on the (poorly constrained) parameters. While our plots show results at a fixed $M_{\rm NS}=1.35\,M_\odot$, they can easily be rescaled to any other choice for the neutron star mass, as the mass predictions only depend on the ratio $M_{\rm NS}/R_{\rm NS}$. We note that at high mass ratios, the choice of $f_{\rm rem}$ has nearly no impact on the constraints. This occurs because the limit on $M_{\rm ej,dyn}$ is more constraining than the limit on $M_{\rm ej,pm}$. At lower mass ratios, on the other hand, $M_{\rm ej,dyn}$ rapidly decreases (it asymptotes to the low values predicted for BNS systems in the near equal-mass regime). In that regime, \ref{fig:pe} shows that the choice of $f_{\rm rem}$ clearly impacts the constraints that we can place on the binary parameters. Conservative upper limits on the BH spin are obtained by choosing $f_{\rm rem}\sim 0.15$. Should more detailed study of post-merger remnants reveal that higher values of $f_{\rm rem}$ are more realistic, our constraints could become noticeably stronger.

We conclude by mentioning three caveats of this analysis. First, as noted above, KN models adopted here assume axial symmetry and a distribution over a $2\pi$ azimuthal angle for the dynamical ejecta. In reality, the dynamical ejecta are predicted to cover only $\sim$ half of the plane and thus $\sim$ half of the orientations in the equatorial plane are expected to be brighter than predicted here. Accounting for the predicted break of symmetry will therefore produce stronger constraints for equatorial viewing angles than those derived here. The second caveat follows from the fact that the composition of the post-merger ejecta in NSBH mergers is uncertain. This is due in large part to the very approximate treatment of neutrinos used in many simulations~\citeNew{WaSe2014,Foucart:2018gis}, but also to the fact that the post-merger ejecta may contain a number of independent components with different geometry, composition, and temperature~\citeNew{KiSe2015,SiMe2017,Fernandez:2018kax}, and the relative contribution of these various components is strongly affected by the unknown strength and large scale structure of the post-merger magnetic field~\citeNew{Christie:2019lim}. Here we adopted a composition intermediate between lanthanide-poor and lanthanide-rich material but note that a different composition would lead to different constraints in the $M_\mathrm{ej,dyn}-M_\mathrm{ej,pm}$ parameter space. For instance, a lanthanide-poor composition for the post-merger ejecta is expected to lead to brighter KNe and thus to result in stronger constraints. Finally, a third caveat is that binaries leading to extremely massive ejecta are not rigorously excluded by our analysis. This is due to the fact that within the grid of models considered here, the more massive ejecta ($M_{\rm dyn}\gtrsim 0.07M_\odot$ and $M_{\rm pm}\gtrsim 0.07M_\odot$) lead to KN that evolve too slowly to pass the observational cuts that we impose on the time evolution of the magnitude of KN, and also because some extreme low-mass systems may have $M_{\rm pm}\geq 0.1M_\odot$, a region not covered by our grid of simulations. The small regions of parameter space untested by this study is shown in \ref{fig:pelim}. We note that on this figure, the excluded region at high NS radii is due to the observational cuts; requiring observations to be sensitive to that region of parameter space may lead to many more false positives. The smaller region at low NS radii and low mass ratio is due to our $M_{\rm pm}< 0.1M_\odot$ limit.

\clearpage
\begin{table*}
\centering
\caption{Follow-up table for all spectroscopically classified transients. Our spectra were obtained with GTC \protect\citeNew{gcn26702, gcn26703}, ePESSTO \protect\citeNew{ScIr2020}, P60+SEDM, and P200+DBSP. The spectroscopic redshifts are listed as well. The objects with a star (*) were first reported to TNS by ALeRCE. Discovery magnitudes reported are extinction-corrected.}

\label{table:S200105ae_spec_followup}
\resizebox{\textwidth}{!}{%
\begin{tabular}{llllllll}
\hline\hline
Name &RA & Dec&TNS & Discov. Mag. &Classification  &Spec. facilities & Spec. Redshift \\ \hline
ZTF20aaertpj  & 14:27:52 &33:34:10 &AT2020pv*& $g$ = 19.88 $\pm$ 0.16 & SN Ib & GTC & 0.026\\ \hline
ZTF20aaervoa  & 15:02:38& 16:28:22 &AT2020pp*&  $g$ = 20.63 $\pm$ 0.30 & SN IIp & GTC & 0.046 \\ \hline
ZTF20aaervyn  & 15:01:27& 20:37:24 &AT2020pq*&  $g$ = 20.62 $\pm$ 0.26 & SN Ia & GTC & 0.112 \\ \hline
ZTF20aaerxsd  & 14:00:54& 45:28:22 &AT2020py &  $g$ = 20.27 $\pm$ 0.23 & SN Ia & GTC & 0.055 \\ \hline
ZTF20aaerqbx  & 15:49:26& 40:49:55 &AT2020ps*&  $g$ = 19.46 $\pm$ 0.15 & SN IIp & GTC & 0.098 \\ \hline
ZTF20aafanxk & 05:35:36  & 11:46:15 & AT2020adk & $r$ = 18.52 $\pm$ 0.25 & SN Ia &P60+SEDM & 0.133\\ \hline
ZTF20aafujqk & 17:57:00 & 10:32:20 & AT2020adg & $r$ = 18.17 $\pm$ 0.10 & SN Ia & P60+SEDM & 0.074\\ \hline
ZTF20aaevbzl & 13:26:41 & 30:52:31 & AT2020adf & $i$ = 19.31 $\pm$ 0.24 & CV & P200+DBSP & 0.0 \\ \hline
ZTF20aafqpum & 03:06:08 & 13:54:48 & SN2020yo & $g$ = 19.76 $\pm$ 0.20 & SN Ia 91-bg & ePESSTO & 0.09\\ \hline
\hline\hline
\end{tabular}}
\end{table*}

\begin{table*}
\centering
\caption{Follow-up table of the candidates identified for S200105ae, reported in Ref.~\protect\citeNew{gcn26810}. The ZTF objects with a star (*) in the TNS column were first reported to TNS by ALeRCE. The spectroscopic (s) or photometric (p) redshifts of the respective host galaxies are listed as well. As a reference, the all-sky averaged distance to the source is $283 \pm 74$\,Mpc, corresponding to a redshift range z = 0.045--0.077.  We use the same rejection criteria described in more detail in section \ref{sec:candidates} here, as follows: slow photometric evolution (slow), hostless, stellar, and slow moving asteroid (asteroid).}

\label{table:S200105ae_followup}
\resizebox{\textwidth}{!}{%
\begin{tabular}{llllllll}
\hline\hline
Name & RA & Dec & TNS & Discov. Mag. & Host/Redshift & rejection criteria \\ \hline
ZTF20aafduvt & 03:36:29  & $-$07:49:35 & AT2020ado & $g$ = 19.57 $\pm$ 0.29 & 0.25 $\pm$ 0.02 (p) & slow \\ \hline
ZTF20aaflndh & 01:22:38  & $-$06:49:34 & AT2020xz  & $g$ = 19.11 $\pm$ 0.11 & 0.091 $\pm$ 0.023 (p) & slow\\ \hline
ZTF20aaexpwt & 06:26:01  & 11:33:39 & AT2020adi & $r$ = 16.95 $\pm$ 0.17 & - & slow \\ \hline
ZTF20aafukgx & 18:23:21 & 17:49:32 & AT2020adj & $r$ = 18.40 $\pm$ 0.15 & - & slow \\ \hline
ZTF20aagijez & 15:04:13 & 27:29:04 & AT2020adm & $r$ = 19.67 $\pm$ 0.3 & 0.061 (s) & slow \\ \hline
ZTF20aagiiik & 16:19:10 & 53:45:38 & AT2020abl*& $g$ = 19.76 $\pm$ 0.22 & 0.13 (s) & slow \\ \hline
ZTF20aafdxkf & 03:42:07  & $-$03:11:39 & AT2020ads & $r$ = 20.02 $\pm$ 0.25 & - & slow \\ \hline
ZTF20aagiipi & 15:33:25 & 42:02:37 & AT2020adl & $g$ = 20.10 $\pm$ 0.32 & 0.39 $\pm$ 0.02 (p) & slow \\ \hline
ZTF20aafsnux & 14:36:01 & 55:11:49 & AT2020dzu & $g$ = 19.67 $\pm$ 0.22 & - & slow \\ \hline
ZTF20aaertil & 14:52:26 & 31:01:19 & AT2020pu* & $g$ = 19.86 $\pm$ 0.18 & 0.093 (s) & slow \\ \hline
ZTF20aafksha & 13:43:54 & 38:25:14 & AT2020adr & $g$ = 20.06 $\pm$ 0.26 & 0.167 (s) & slow \\ \hline
ZTF20aagjemb & 14:51:26 & 45:20:41 & AT2020adh & $r$ = 20.90 $\pm$ 0.02 & 0.21 $\pm$ 0.06 (p) & slow \\ \hline
ZTF20aafefxe & 07:47:24  & 14:42:24 & AT2020adt & $g$ = 21.0 $\pm$ 0.18 & 0.09 $\pm$ 0.05 (p) & slow \\ \hline
ZTF20aafaoki & 05:13:14  & 05:09:56  & AT2020adq & $r$ = 19.21 $\pm$ 0.28 & - & slow \\ \hline
ZTF20aafexle & 04:20:31  & $-$09:30:28 & AT2020adn & $r$ = 19.67 $\pm$ 0.30 & 0.18 $\pm$ 0.02 (p) & stellar \\ \hline
ZTF20aaegqfp & 07:49:02 & 12:29:26 & AT2020dzt & $r$ = 19.37 $\pm$ 0.27 & - & asteroid \\ \hline
\hline\hline

\end{tabular}}
\end{table*}

\begin{table*}
\centering
\caption{Follow-up table of the candidates identified for S200115j, reported in Ref.~\protect\citeNew{gcn26767}. As a reference, the all-sky averaged distance to the source is $340 \pm 79$\, Mpc, corresponding to a redshift range z = 0.056--0.089.}

\label{table:S200115j_followup}
\resizebox{\textwidth}{!}{%
\begin{tabular}{llllllll}
\hline\hline
Name & RA & Dec & TNS & Discov. Mag. & Host/Redshift  & rejection criteria \\ \hline
ZTF20aahenrt & 09:32:53 & 72:23:06 & AT2020axb & $g$ = 20.55 $\pm$ 0.29 & 0.16 $\pm$ 0.04 (p)& slow \\ \hline
ZTF20aagjqxg & 02:59:39 & 06:41:11 & AT2020aeo & $g$ = 20.65 $\pm$ 0.26 & - & slow \\ \hline
ZTF20aahakkp & 05:07:55 & 56:27:50 & AT2020bbk & $g$ = 15.67 $\pm$ 0.08 & - & slow \\ \hline
ZTF20aafqulk & 03:39:45 & 27:44:05 & AT2020yp & $g$ = 20.74 $\pm$ 0.21 & - & stellar \\ \hline
ZTF20aafqvyc & 03:47:58 & 38:26:32 & AT2020yq & $r$ = 20.39 $\pm$ 0.19 & - & asteroid \\ \hline
\hline\hline
\end{tabular}}
\end{table*}

\end{supplement}

\bibliographystyleNew{naturemag}
\bibliographyNew{references}

\end{document}